\documentclass[a4paper,twocolumn,11pt,accepted=2021-10-27]{quantumarticle}
\pdfoutput=1
\usepackage{cite}
\usepackage{diagbox}
\usepackage{gnuplottex}
\usepackage{mathtools} 
\usepackage{float}
\usepackage{array}
\usepackage[english]{babel}
\usepackage[utf8]{inputenc}
\usepackage[T1]{fontenc}
\usepackage{amsmath}
\usepackage[table]{xcolor}
\usepackage{amsfonts}
\usepackage{dsfont}
\usepackage{algorithm}
\usepackage[noend]{algpseudocode}
\usepackage{bbold}
\usepackage{mathtools}
\usepackage{amssymb}
\usepackage{bm}
\usepackage{bbm}
\usepackage{tabularx, booktabs}
\usepackage{xspace}
\usepackage{listings}
\usepackage{soul}
\newcolumntype{Y}{>{\centering\arraybackslash}X}
\usepackage{braket}
\definecolor{dgreen}{rgb}{0,.5,0}
\definecolor{dblue}{rgb}{0,0,.5}
\definecolor{dred}{rgb}{0.5,0,.5}
\usepackage{bm}
\usepackage{scalerel,stackengine}
\stackMath
\newcommand\reallywidehat[1]{%
\savestack{\tmpbox}{\stretchto{%
  \scaleto{%
    \scalerel*[\widthof{\ensuremath{#1}}]{\kern-.6pt\bigwedge\kern-.6pt}%
    {\rule[-\textheight/2]{1ex}{\textheight}}
  }{\textheight}%
}{0.5ex}}%
\stackon[1pt]{#1}{\tmpbox}%
}
\newcommand{\bruno}[1]{\textcolor{red}{\textbf{\underline{Bruno :}} #1}}

\begin{document}


\title{Encoding strongly-correlated many-boson wavefunctions on a photonic quantum computer: application to the attractive Bose-Hubbard model} 

\author{Saad Yalouz}
\email{yalouzsaad@gmail.com}
\affiliation{Laboratoire de Chimie Quantique, 
CNRS/Université de Strasbourg, 4 rue Blaise Pascal, 67000 Strasbourg, France}
\affiliation{Theoretical Chemistry, Vrije Universiteit, De Boelelaan 1083, NL-1081 HV, Amsterdam, The Netherlands}
\affiliation{Instituut-Lorentz, Universiteit Leiden, P.O. Box 9506, 2300 RA Leiden, The Netherlands}
\author{Bruno Senjean}
\email{bruno.senjean@umontpellier.fr}
\affiliation{ICGM, Univ Montpellier, CNRS, ENSCM, Montpellier, France}
\affiliation{Instituut-Lorentz, Universiteit Leiden, P.O. Box 9506, 2300 RA Leiden, The Netherlands}

\author{Filippo Miatto}
\email{filippo@xanadu.ai}
\affiliation{Xanadu, Toronto, ON, M5G 2C8, Canada}

\author{Vedran Dunjko}
\email{v.dunjko@liacs.leidenuniv.nl}
\affiliation{Universiteit Leiden, P.O. Box 9506, 2300 RA Leiden, The Netherlands}

\begin{abstract}
Variational quantum algorithms (VQA) are considered as some of the most promising methods to determine the properties of complex strongly correlated quantum many-body systems, especially from the perspective of devices available in the near term.
In this context, the development of efficient quantum circuit ansatze to encode a many-body wavefunction is one of the keys for the success of a VQA. 
Great efforts have been invested to study the potential of current quantum devices to encode the eigenstates of fermionic systems,
but little is known about the encoding of bosonic systems. 
In this work, we investigate the encoding of the ground state of the (simple but rich) attractive Bose--Hubbard model using a Continuous-Variable (CV) photonic-based quantum circuit.
We introduce two different ansatz architectures and demonstrate that the proposed continuous variable quantum circuits can accurately encode (with a fidelity higher than 99\%) the strongly correlated many-boson wavefunction with just a few layers, in all many-body regimes and for different number of bosons and initial states.
Beyond the study of the suitability of the ansatz to approximate the ground states of many-boson systems, we also perform initial evaluations of the use of the ansatz in a variational quantum eigensolver algorithm to find it through energy minimization. 
To this end we also introduce a scheme to measure the Hamiltonian energy in an experimental system, and study the effect of sampling noise.
\end{abstract}

\maketitle

\section{Introduction}

In recent years, great efforts have been deployed to study the potential of noisy intermediate-scale quantum (NISQ) computers to tackle tasks that are hard to treat for classical computers. 
One of the most emblematic research lines in the field is the development of near-term algorithms to determine the properties of complex quantum many-body systems. 
Variational quantum algorithms (VQA) have shown to be very efficient and promising for encoding complex wavefunctions of various kinds of systems. 
Applications of these algorithms range from molecular systems in quantum chemistry~\cite{peruzzo2014variational, omalley2016scalable,shen2017quantum,kandala2017hardware,hempel2018quantum,colless2018computation,mazzola2019nonunitary,nam2020ground,obrien2019calculating,sokolov2021microcanonical,arute2020hartree}, interacting-spins and electrons models from condensed matter~\cite{cade2020strategies,fujii2020deep,xu2020test} or vibrational Hamiltonians~\cite{ollitrault2020hardware}, to cite but a few.

{
In practice, the success of a VQA in encoding a many-body wavefunction depends on the properties of the associated quantum circuit, or ``ansatz'', that is used.  
Ideally, to satisfy the needs and limitations of the NISQ era, an ansatz should be expressive (i.e.~for example, having the ability to generate accurate ground-state wavefunctions or energies), resource efficient (i.e.~involving as few gates/layers as possible) and trainable (i.e.~inducing optimization landscapes amenable to classical optimization algorithms).
In the case of qubit-based quantum circuits, the above properties of ansatze for VQAs have been  studied extensively:
examples are the unitary coupled-cluster ansatz~\cite{romero2021variational} introduced for electronic structure problems, adaptive methods designed to produce expressive ansatze at lower complexity cost~\cite{grimsley2019adaptive,tang2019qubit}, and different strategies to solve condensed matter models with strong quantum correlation~\cite{cade2020strategies,montanaro2020compressed}.
The great majority of these works has been essentially devoted to the case of many-fermion systems.
}
{In contrast, little attention has been paid to study the potential of current devices to produce interesting ansatze for the study of strongly interacting many-boson systems.
On the classical computing side, state-of-the-art methods such as quantum Monte-Carlo (QMC) simulations have proven to be  effective~\cite{masaki2019quantum,krauth1996quantum,del2010interacting,purwanto2004quantum,pollet2013review, bogner2019variational,lkacki2016locating,kawaki2017phase,dogra2016phase,lv2014two,rossi2009exact}, as they do not suffer from the infamous sign problem of fermionic systems.
However, QMC is usually restricted to ground-state calculations, although extensions to molecular properties~\cite{assaraf2000computing,filippi2000correlated,sorella2010algorithmic,assaraf2003zero,chiesa2005accurate} and excited states~\cite{ceperley1988calculation,zimmerman2009excited,williamson1998diffusion,grimes1986quantum,feldt2020excited} have been developed.
The rise of quantum computing can open alternative routes to ground-state calculations (and beyond) for many-boson systems,
for instance based on superconducting circuits~\cite{gangat2013deterministic,wilkinson2020superconducting,Roushan2017spectroscopy,hacohen2015cooling,deng2016superconducting}, cold-atom lattices~\cite{Cotler2019Quant,kyriienko2020quantum} or photonic quantum devices~\cite{huh2015boson,huh2017vibronic,sparrow2018simulating,quesada2019franck,jahangiri2020quantum,jahangiri2021quantum,kalajdzievski2018continuous,zhong2020quantum} (such as Gaussian boson sampling experiments~\cite{huh2015boson,huh2017vibronic,jahangiri2020quantum,jahangiri2021quantum}), essentially used as analog quantum simulators.
Indeed, many extensions of VQAs have been proposed recently to compute molecular and excited-state properties~\cite{kassal2009quantum,obrien2019calculating,mitarai2020theory,parrish2019hybrid,sokolov2020microcanonical,mcclean2017hybrid,ollitrault2019quantum,nakanishi2019subspace,ibe2020calculating,lee2018generalized,higgott2019variational,jones2019variational,jouzdani2019method,parrish2019quantum,bauman2019quantum,motta2020determining,zhang2020variational,yalouz2021state}, and one
can expect similar developments for many-boson systems in the near future.}

{
Motivated by the development of VQAs, we investigate in this work the use of digital photonic-based continuous variable (CV) quantum computers~\cite{arrazola2021quantum,weedbrook2012gaussian,bromley2020applications,pfister2019continuous} to encode strongly correlated many-boson wavefunctions.} In practice, CV devices are digital quantum computers that can manipulate information by entangling and measuring quantum states of light using various types of quantum gates. 
The bosonic nature of such photonic devices leads to a one-to-one mapping between the photonic modes of a CV computer and the bosonic modes of other many-boson systems.
Such a mapping was first employed in Ref.~\cite{huh2015boson}
to generate molecular vibronic spectra, where the $N$ photons in $M$ optical modes can simulate $N$ molecular vibrational quanta in $M$ vibrational modes.
Bosonic modes of CV devices were also used to efficiently encode the field degrees of freedom in a wavelet basis in quantum field theories~\cite{marshall2015quantum,brennen2015multiscale}.
For variational algorithms, the CV setting is then particularly well-suited to explore new types of quantum circuit ansatz to encode strongly correlated many-boson wavefunctions.
The rich complexity of such type of wavefunctions
can be qualitatively reproduced by studying the attractive Bose--Hubbard (BH) model.
The ground state of this model has attracted a lot of attention recently due to its rich transitional behavior that evolves between the so-called Schr\"odinger cat-like states to superfluid regimes~\cite{kuhner1998phases,kuhner2000one,jack2005bose,oelkers2007ground,julia2010macroscopic,mele2011improved,mansikkamaki2021phases}.
Furthermore, the BH model has been successfully employed to describe the physics of ultracold atoms and molecules in optical lattices~\cite{theis2004tuning,goral2002quantum,buchler2003supersolid,bloch2005ultracold,bloch2008many,baier2016extended}, quantum cooling protocols~\cite{hacohen2015cooling}, the evolution of vibrational bosons in $\alpha$-helices proteins~\cite{pouthier2003two,edler2004direct}
and the exotic phases of helium-4~\cite{matsuda1970off,liu1973quantum}.

{
In this paper, we study the attractive BH model as an interesting system and focus on two aspects of the aforementioned critical properties of ansatze for VQA applications: the expressibility and the resource efficiency to encode the ground-state wavefunction of the system on CV devices.
Note that, in analogy with the qubit cases, the implementation costs of a photonic circuit depend on numerous factors.
Obviously, they depend on depth and total gate number,
but not all gates are equally costly (e.g. linear optical elements are much cheaper than gates involving non-linear terms), nor they introduce an equal amount of noise and losses. 
We are thus facing a multi-objective optimization problem as, for instance, using cheaper gates may come at the expense of greater depth to achieve the same theoretical capacity to represent ground states accurately.
Further, in the light of near-term implementations, it makes additional sense to study ansatz tailored to the actual experiments performed in the labs or to those available in the near future.
To elucidate the issues of trade-offs between various resource requirements and expressibility, we propose and analyze two CV quantum circuit architectures tailored to near-future experimental set-ups that can make use of single-mode non-linear phase shifts such as the Kerr gate: one designed to minimize the total gate count and the other designed to minimize the total circuit depth.
By performing VQAs with the state-infidelity as a cost function, we show that the many-boson ground state of the BH model can be successfully encoded on both CV quantum circuits with a high fidelity and a relatively small number of gates and parameters.
In practice, the state-infidelity cannot be computed for large systems (as it requires the knowledge of the exact wavefunction). A realistic experiment on a true quantum device would require the ability to measure on a CV circuit the ground-state energy instead.
Thus, we propose a photon-counting protocol to measure the expectation value of the BH Hamiltonian in order to simulate a variational quantum eigensolver (VQE) algorithm, as it would be done in a true experiment, with and without sampling noise.
}

The paper is organized as follows. 
After a brief introduction to the attractive BH Hamiltonian and the exact diagonalization method in sections~\ref{subsec:ham} and ~\ref{subsec:exact}, we illustrate the exotic structural transition of the BH ground state on three small-sized networks in section~\ref{subsec:illustration}.
After introducing the context of VQAs in section~\ref{subsec:VQA}, 
we present the architectures of our proposed ansatze in section~\ref{subsec:ansatze}. 
Numerical investigations of their encoding ability
are performed in section~\ref{sec:num_results_expressibility}.  
We extend our study to ideal and realistic simulations of a VQE algorithm in section~\ref{sec:VQE}.
Finally, conclusions and perspectives are provided in section~\ref{sec:conclusion}.



\section{Bose-Hubbard model}\label{sec:BH}

In this first section, we present a general introduction of the attractive BH model. 
Our objective is to give an overview of the properties of the model and to illustrate the rich behaviors that arise from the increase of the many-body interaction. 
We will focus on three small-sized BH models, namely the BH dimer and the three- and four-site periodic chains. 
These systems will be used later in this paper as test-bed for the quantum ansatze designed herein to encode many-boson wavefunctions on photonic devices.

\subsection{Hamiltonian and regimes of interaction}\label{subsec:ham}

The attractive BH model describes the physics of $N_B$ bosons hopping between $N_S$ sites of a network with local attractive many-body interactions. The associated Hamiltonian operator reads
\begin{equation}\label{eq:HBH}
    \hat{\mathcal{H}}  =   -J \sum_{\langle p,q\rangle}^{N_S}  (a_p^\dagger a_q + a_q^\dagger a_p )  - \frac{U}{2}\sum_p^{N_S}  n_p(n_p - 1),
\end{equation}
where $J>0$ is the hopping parameter describing the delocalization of the bosons throughout the connected nodes of the BH network, and $U>0$ is the many-body interaction term corresponding to the local boson-boson interaction that occurs when at least two particles occupy the same site. 
In this work, $J$ is the energy unit (i.e.~we set $J\equiv 1$ in every simulation).


To characterize the competition between the many-body interactions and the hopping terms in the system, a dimensionless parameter
\begin{equation}
    \Lambda = \frac{N_BU}{J}
\end{equation} 
is commonly introduced.
Typically, $\Lambda \ll 1 $ represents a regime of weak interaction (the so-called ``Super-fluid regime'') in which the physics of the system is governed by the hopping terms making the bosons able to efficiently delocalize over the network.
The case $\Lambda \sim 1 $ corresponds to an intermediate quantum regime where hoppings and local many-particle interactions compete together and where phase transitions usually occur.
The last case, $\Lambda \gg 1 $, corresponds to the  regime where the many-body interaction dominates (the so-called ``Fock regime'').
In the case of attractive interactions, the resulting state tends to localize the bosons on local sites of the system (giving rise to macroscopic-cat states as illustrated later in this work).

\subsection{Ground state reference: exact diagonalization}\label{subsec:exact}
 
To investigate the exotic phases encoded in the attractive BH model, we compute the ground state of the Hamiltonian $\hat{\mathcal{H}}$ exactly by solving the time-independent Schr\"{o}dinger equation
\begin{equation}
    \hat{\mathcal{H}} \ket{\Psi_0} =  E_0 \ket{\Psi_0},
    \label{eq:GS}
\end{equation}
where $\ket{\Psi_0}$ is the ground state of the system with the associated energy $E_0$. 
We solve Eq.~(\ref{eq:GS}) by exact diagonalization (ED) to obtain a good reference state to compare with our quantum ansatz developed for CV devices.
Note that the way ED works in theory  may appear simple: given a system of $N_B$ bosons evolving on a network of $N_S$ sites, one builds an exact matrix representation of the associated Hamiltonian and diagonalize it. But beyond its apparent simplicity, the method requires in practice some technical efforts of implementations (see Refs.~\cite{raventos2017cold,zhang2010exact} for an introduction of the numerical approach). 
In the case of the attractive BH model, the dimension $D_{\hat{\mathcal{H}}}$ of the resulting matrix is equal to the dimension of the Fock space of the system,
\begin{equation}
    D_{\hat{\mathcal{H}}} = \frac{(N_B + N_S-1)!}{N_B! (N_S-1)!},
\end{equation}
which represents the number of ways of distributing $N_B$ bosons on $N_S$ sites. 
See Tab.~\ref{table:dimensions}, for an overview of the scaling of the $D_{\hat{\mathcal{H}}} $ as a function of $N_S$ and $N_B$.

While exact diagonalization provides access to the exact solution of the problem, the exponential scaling of the Fock space dimension with respect to the number of sites and bosons limits the use of ED to very small systems in practice.

\begin{table} 
\small
\centering
\begin{tabular}{|c||l|l|l|l|l|l|} 
\hline
\backslashbox[4.5em]{ $\bm{N_S}$ }{ $\bm{N_B}$ }  & \makebox[1.5em]{\textbf{2}}  & \makebox[1.5em]{\textbf{3}}   & \makebox[1.5em]{\textbf{4}}   & \makebox[1.5em]{\textbf{5}}  & \makebox[1.5em]{\textbf{8}}    & \makebox[1.5em]{\textbf{16}}      \\ 
\hline\hline
\textbf{2} & 3  & 4   & 5 & 6  & 9    & 17      \\ 
\hline
\textbf{3} & 6  & 10  & 15 & 21 & 45   & 153     \\ 
\hline
\textbf{4} & 10 & 20  & 35 & 56  & 165  & 969     \\ 
\hline
\textbf{8} & 36 & 120 & 330 & 792  & 6435 & 245157  \\
\hline
\end{tabular}
\caption{Dimension $D_{\hat{\mathcal{H}}}$ of the Fock space of the BH model as a function of the number of sites $N_S$ and bosons $N_B$. }
\label{table:dimensions}
\end{table}

\subsection{Illustration of the regimes of interaction: exact diagonalization on small-sized systems}\label{subsec:illustration}

As the many-body interaction increases, the ground state of the attractive BH model exhibits rich transitional behaviors.
Such transitions have already been extensively discussed in several recent works (see for example~\cite{julia2010macroscopic,oelkers2007ground,mansikkamaki2021phases}). 
Here, we briefly illustrate some of these exotic phases in three systems (that will be treated later using quantum ansatze developed on a CV device), namely the BH dimer, the three-site and the four-site periodic chain. To do so, we employ the Inverse Participation Ratio (IPR) \cite{bauer2013area,hopjan2020many,giraud2009entropy,giraud2007entanglement} and the von Neumann entropy~\cite{lukin2019probing, islam2015measuring, beugeling2015global}
as two indicators to follow the expansion of $\ket{\Psi_0}$ in the configuration basis when $\Lambda$ increases. 
Applied to the Fock state basis, the IPR estimates the number of different bosonic configurations forming $\ket{\Psi_0}$ by linear combination, and reads
\begin{equation}
    \text{IPR}(\ket{\Psi_0}) =   \Big( \sum_{n_1,\ldots,n_{N_S}}|\braket{n_1,\ldots,n_{N_S} | \Psi_0}|^4\Big)^{-1},
\end{equation}
where $\ket{n_1,\ldots,n_{N_S}}$ is a Fock state with $n_q$ bosons on site $q$ (and $\sum_q n_q = N_B$).
Within this definition, a state $\ket{\Psi_0}$ localized on a single state of the Fock-state basis has an IPR of 1. 
By contrast, the IPR of a state uniformly
delocalized over all the Fock states is equal to $D_{\hat{\mathcal{H}}}$. The von Neumann entropy is a measure used to quantify the entanglement (non-local quantum correlation) existing between different sub-parts of a quantum system.
In many-body systems, it quantifies the degree of entanglement and correlation arising between the subsystem $p$ and all the other modes (also commonly called ``bi-partite entanglement'').
It reads
\begin{equation}\label{eq:S}
    S = - \sum_k \lambda_k \log(\lambda_k ),
\end{equation}
where $\lambda_k$ are the eigenvalues
of the reduced density matrix of mode $p$,
\begin{equation}
    \hat{\sigma}_p = \sum_{n_k = 0}^{N_B} \lambda_k \ket{n_k}\bra{n_k},
\end{equation}
defined by taking the partial trace of the full-system density operator $\hat{\rho}$ over the Fock-state basis of all the other modes in the systems, $\hat{\sigma}_p = \mathbf{Tr}_{q\neq p}[\hat{\rho}]$ (note that in our case, the full density operator is $\hat{\rho} = \ket{\Psi_0}\bra{\Psi_0}$ as one focuses on the ground state).
For the sake of simplicity, we will only focus on a single-site von-Neumann entropy as defined in Eq.~(\ref{eq:S}).

 \begin{figure}[h!]
    \centering 
    \includegraphics[width=\columnwidth]{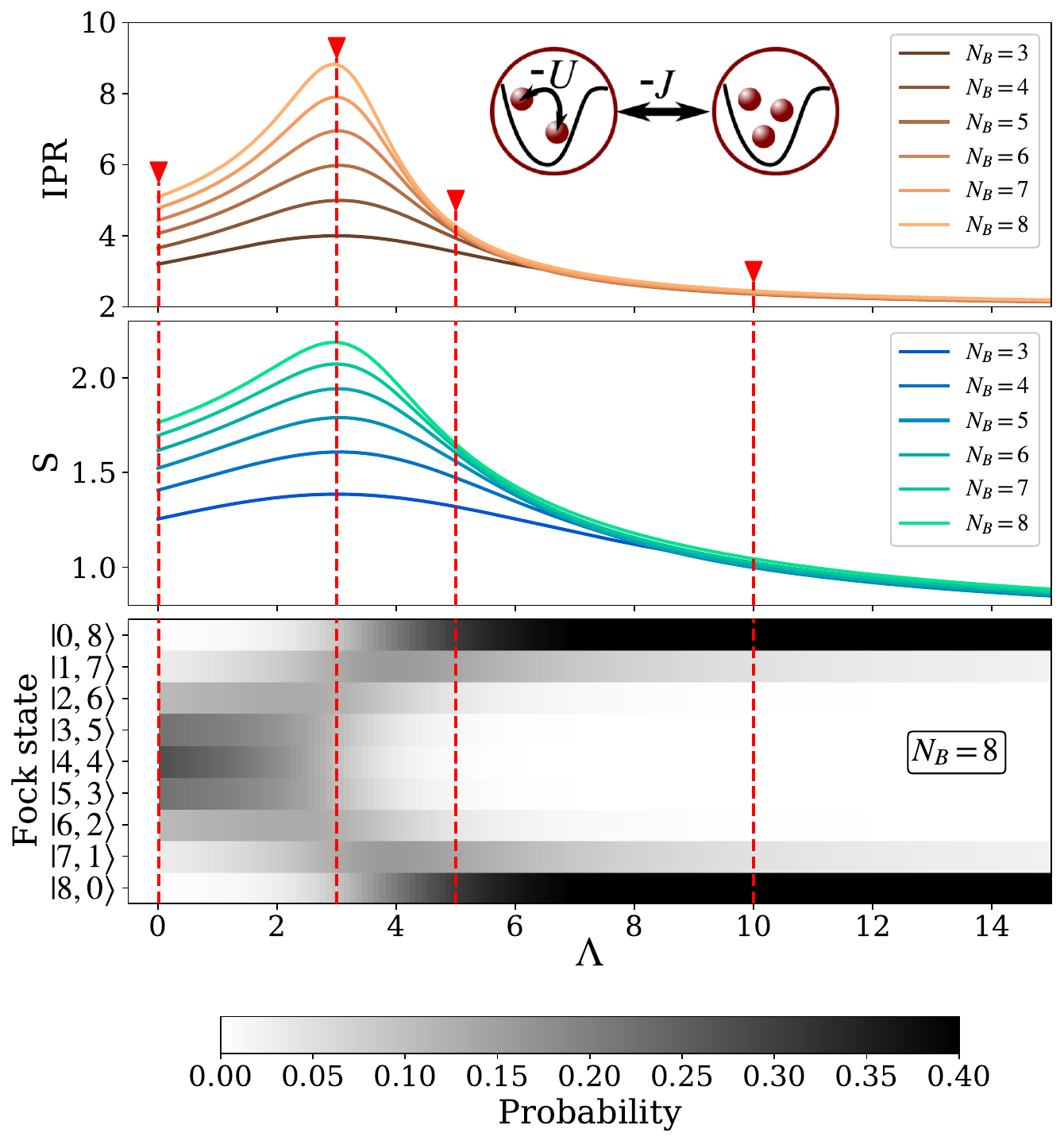}
    \caption{ \textbf{Ground-state properties of the BH dimer.} Evolution of the ground-state structure as a function of $\Lambda$ (obtained with ED). \textbf{Upper panel:} IPR values for different total number of bosons $N_B=3 ,\ldots,8$. 
    \textbf{Middle panel:} Single site entropy $S$ values for different values of the total number of bosons $N_B=3,\ldots,8$.
    \textbf{Lower panel:} Occupancy probability of the ground state in the Fock state basis for $N_B=8$. Probabilities are illustrated with shades of grey, ranging from white for zero to black for $40\%$. In all panels, the vertical dashed lines mark values of $\Lambda$ representative of three different many-body regimes. These values will be used later on to illustrate the expressibility of our quantum ansatze on a CV circuit. }
    \label{fig:dimer_GS}
\end{figure}

\begin{figure}[t]
    \centering
    \includegraphics[width=\columnwidth]{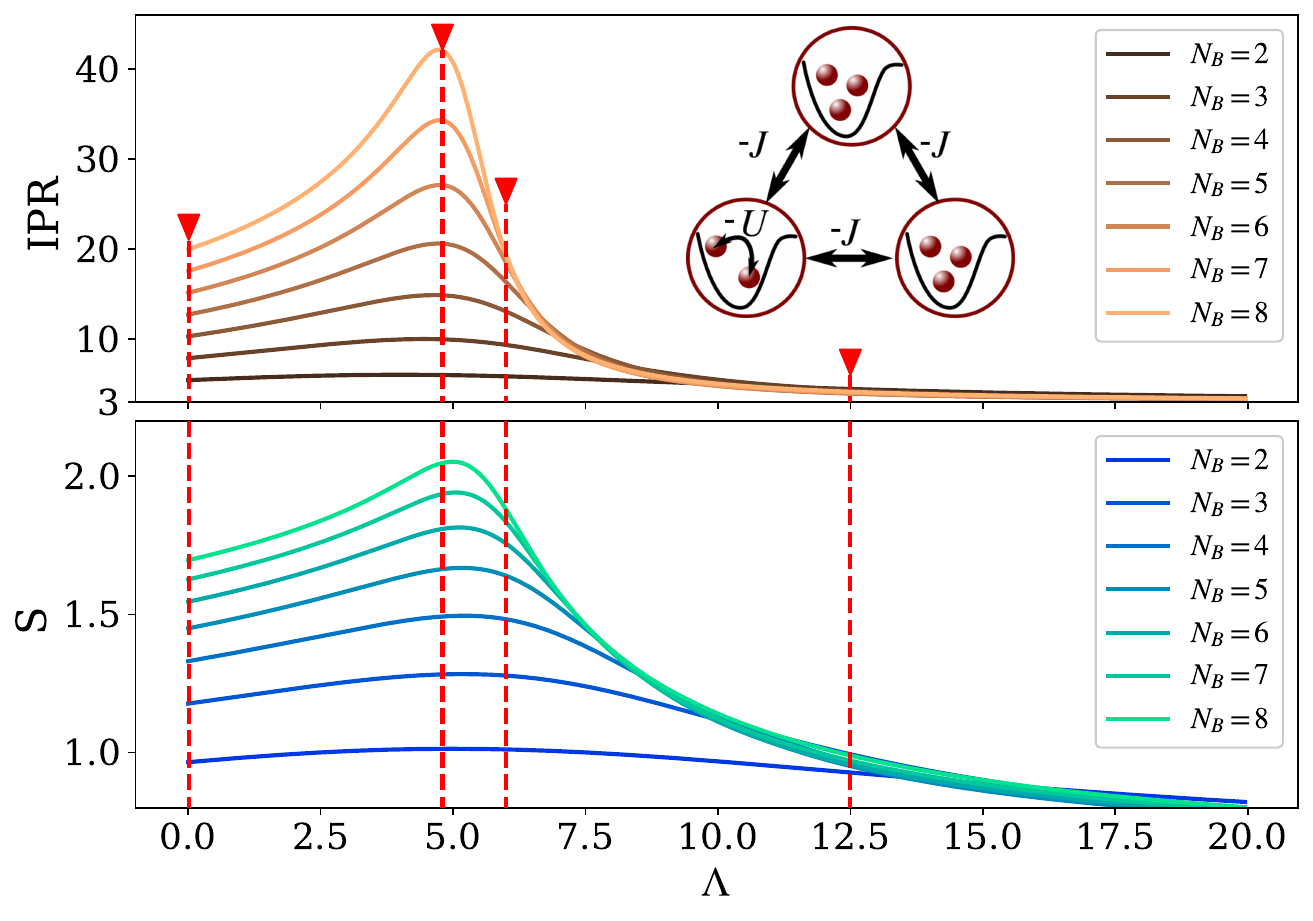} 
    \includegraphics[width=\columnwidth]{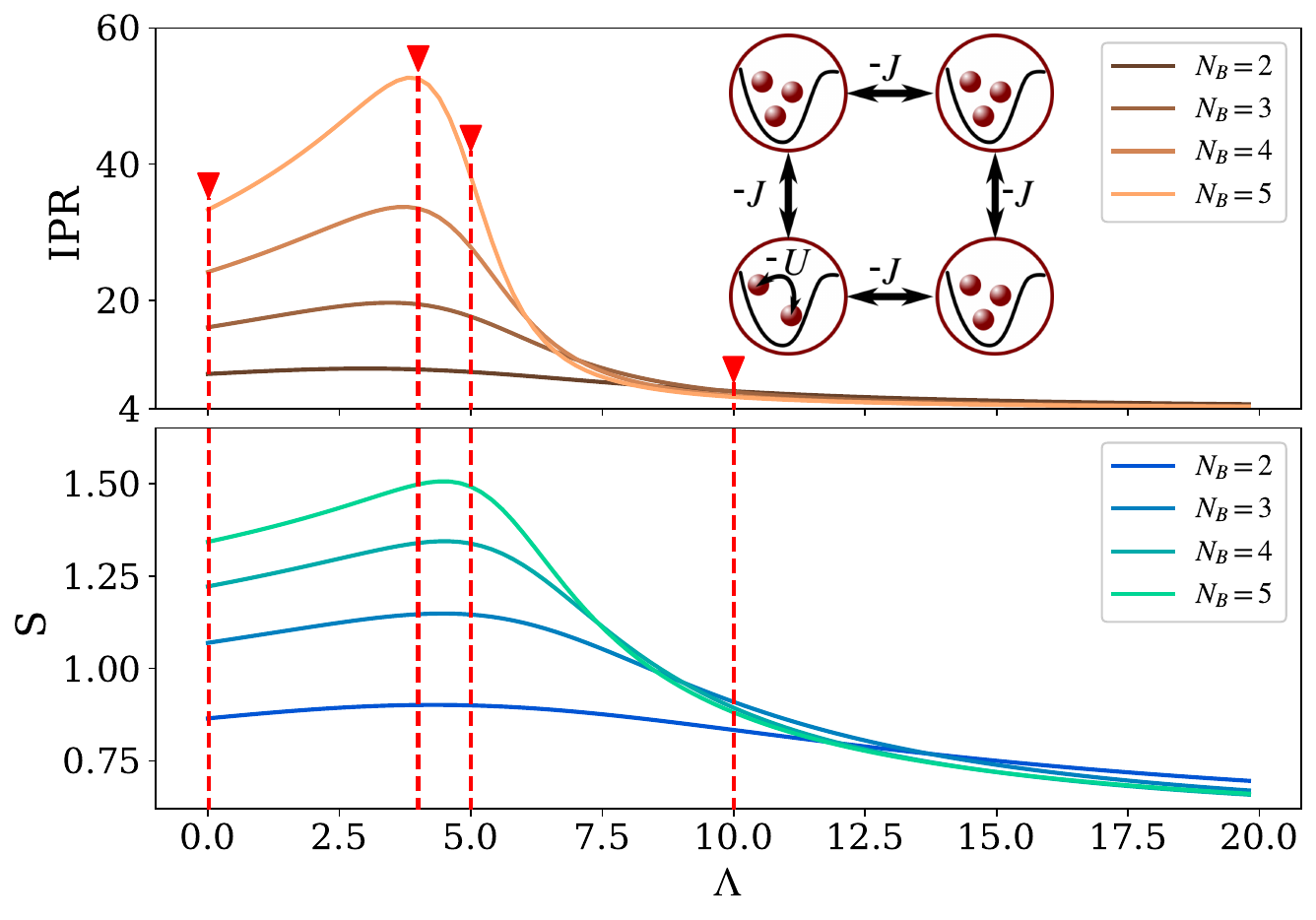} 
    \caption{
    { \textbf{Ground-state properties of the three- and four-site BH periodic chains.}
    Evolution of the IPR and the single-site entropy $S$ of the ground state as a function of $N_B$ and $\Lambda$ (obtained by ED). \textbf{Upper panel:} Three-site BH periodic chain. \textbf{Lower panel:} Four-site BH periodic chain. In both panels, vertical dashed lines mark values of $\Lambda$ representative of three different many-body regimes. These values will be used later on to illustrate the expressibility of our quantum ansatze. 
    }
    } \label{fig:3_4sites}
\end{figure}

In Figure \ref{fig:dimer_GS}, we show the evolution of the ground state properties of the BH dimer as a function of the interaction parameter $\Lambda$ (for different fixed number of bosons $N_B$). 
When $\Lambda \sim 0$, the system is in a superfluid phase: every bosons propagate independently on the dimer and do no interact together. 
The independent-boson picture leads to a product-state solution for the ground state as $\ket{\Psi_0} = \frac{1}{\sqrt{2}\sqrt{N_B!}}(a_1^\dagger + a_2^\dagger)^{N_B}\ket{0,0}$ (with $\ket{0,0}$ being the vacuum state of the dimer). 
Then, increasing $\Lambda$ leads to a progressive restructuring of the ground state as readily seen by the increasing IPR.
This shows a spreading of the state over the whole Fock basis that is also accompanied by an increasing single-site von-Neumann entropy. 
Both quantities then reach a maximum value for $\Lambda \approx 3$. 
The large variations of these two measures are not without significance and reveal the appearance of quantum correlations and entanglement in the system. 
Indeed, in the region $2\leq \Lambda \leq 4$ the shape of the ground state gets more complex as it involves a larger number of bosonic configurations (actually almost all the Fock states are involved), and locally the bosonic modes get strongly entangled.
Finally, as $\Lambda$ increases the structure of the ground state loses in complexity and progressively tends to the well-known ``cat-like'' form~\cite{julia2010macroscopic,oelkers2007ground} representing a superposition of two macroscopically occupied states (i.e.~ $\ket{\Psi_0} = (\ket{N_B,0} + \ket{0,N_B} )/\sqrt{2}$). 
The IPR reaches a value of 2, thus indicating the importance of these two fully localized bosonic configurations. 
The entanglement entropy $S$ of the modes decreases as less bosonic configurations get correlated.
To better picture the full restructuring process occurring during this exotic quantum transition, 
we represent the decomposition of the BH dimer's ground state in the Fock basis as a function of $\Lambda$ in the lower panel in Fig.~\ref{fig:dimer_GS}.


Turning to larger systems, similar exotic quantum transitions can be observed. 
In figure~\ref{fig:3_4sites} we see that the IPR and the single-site von-Neumann entropy also reach their maximal value around $\Lambda \approx 5$ for the three- and the four-site periodic chains.
In this regime, the complexity of the ground state significantly grows with the total number of sites $N_S$ but also with the total number of bosons $N_B$. 
As an illustration, there are around $\text{IPR}^{max} \approx 42$ bosonic configurations involved in the structure of the ground state at the maximum value of the IPR ($\Lambda \approx 4.8$) for the three-site BH model with $N_B=8$. 
This number gets even higher for the four-site model with $N_B=5$ and $\Lambda \approx 4$ with a total number of $\text{IPR}^{max} \approx 55$ bosonic configurations. 
When $\Lambda$ goes beyond this region of high complexity, we retrieve the so-called cat-states
as indicated by the number of macroscopically-occupied states ($\text{IPR}=3$ and $4$ for the three and four-site systems, respectively, e.g.~for the three-site model we have $\ket{\Psi_0} = (\ket{N_B,0,0}+\ket{0,N_B,0}+\ket{0,0,N_B})/\sqrt{3}$).

\section{Photonic-gate-based ansatze}\label{sec:ansatz}
 
In the following, we introduce the different theoretical ingredients to encode the BH ground state on a photonic-gate-based device.

\subsection{Variational Quantum Algorithms}\label{subsec:VQA}
  
For the study of many-body systems in the NISQ era, VQAs represent state-of-the-art methods  that have been proficiently used to determine ground (and excited) states' properties of various systems (see for example Refs.~\cite{romero2021variational,grimsley2019adaptive,tang2019qubit,cade2020strategies,montanaro2020compressed} and the recent review in Ref.~\cite{cerezo2020variational}). 
The protocol of a VQA usually follows the steps: $i$) map the problem onto the quantum device, $ii$) use a quantum circuit as an ``ansatz'' to generate a trial state that encodes the wavefunction of the many-body system,
$iii$) evaluate a given cost function by repeated measurements,
$iv$) optimize the quantum gates parameters of the ansatz circuit to minimize the cost function governed by a variational principle.
In practice, one applies a unitary transformation $\hat{U}(\vec{\theta})$ with tunable gate parameters (formally represented by the vector $\vec{\theta}$) to transform an initial state $\ket{\Psi_{ini}}$ into a trial state 
\begin{equation}\label{eq:transfo}
    \ket{ \Psi(\vec{\theta}) } = \hat{U}(\vec{\theta})\ket{\Psi_{ini}}.
\end{equation} 
Often $\ket{\Psi_{ini}}$ is chosen based on some physical motivations which depends on the targeted problem (e.g.~such as the Hartree--Fock state in electronic structure problems).
If one seeks the ground state of a system, the cost function is usually the energy of the trial state defined as
\begin{equation}
   E(\vec{\theta}) = \bra{ \Psi(\vec{\theta}) }  \hat{\mathcal{H}} \ket{ \Psi(\vec{\theta}) } \geq E_0,
\end{equation}
where $E_0$ is the exact ground-state energy which represents a natural lower bound for the energy of the trial state $\ket{\Psi(\vec{\theta})}$.
In a real VQA experiment, the expectation value $E(\vec{\theta})$ 
is estimated by statistical sampling.
The energy estimator will have an uncertainty based on the number of state preparations and measurements used. 
This energy is then passed to an external classical optimization routine which provides a new set of parameters $\vec{\theta}$ used to update the quantum circuit.
The whole optimization process is repeated until global convergence is reached (i.e.~until $E(\vec{\theta})$ is minimized).

In the NISQ era, the overall performance of a VQA in encoding a many-body wavefunction is highly dependent on the properties of the ansatz that is considered (as mentioned in the introduction: expressibility, resource efficiency and trainability).
In our study of CV ansatz, our attention turns towards the expressibility and resource efficiency,
while the investigation of the trainability is left for future work.
However, these properties are not independent,
and the final performance should include the specification of the optimization as it may heavily influence
the expressibility and resource efficiency of the VQA to achieve a given precision of the targeted state or energy.
For this reason, studying all three properties 
is very difficult.

In the following, we introduce two different ansatze architectures and we apply them to the three BH systems introduced before. 
The different architectures address (to some extent) the question of expressibility and resource efficiency. For this, we characterize the capability of the associated circuit unitaries $\hat{U}(\vec{\theta})$
(see Eq.~\eqref{eq:transfo}) to accurately map different initial states $\ket{\Psi_{ini}}$ to the exact ground state $\ket{\Psi_0}$ of the three systems.
Note that only a few works have tried to investigate the expressibility of ansatze for the case of many-body systems (always with a special emphasis for qubit devices and many-electron systems, e.g.~the study of the unitary coupled-cluster-ansatz in \cite{evangelista2019exact}). As a results, a lot of attention has been directed to many-fermion systems (see for example~\cite{grimsley2019adaptive,tang2019qubit,cade2020strategies,montanaro2020compressed,sokolov2020quantum, mizukami2020orbital, yalouz2021state}),
but little work has been done on quantum ansatze to encode many-boson wavefunctions (for a notable exception, see the implementation of the unitary vibrational-coupled-cluster ansatz introduced in Ref.~\cite{ollitrault2020hardware}).

\subsection{Photonic quantum circuits as ansatze for many-boson wavefunction}\label{subsec:ansatze}

\begin{figure}[h]
    \centering
    \includegraphics[width=\columnwidth]{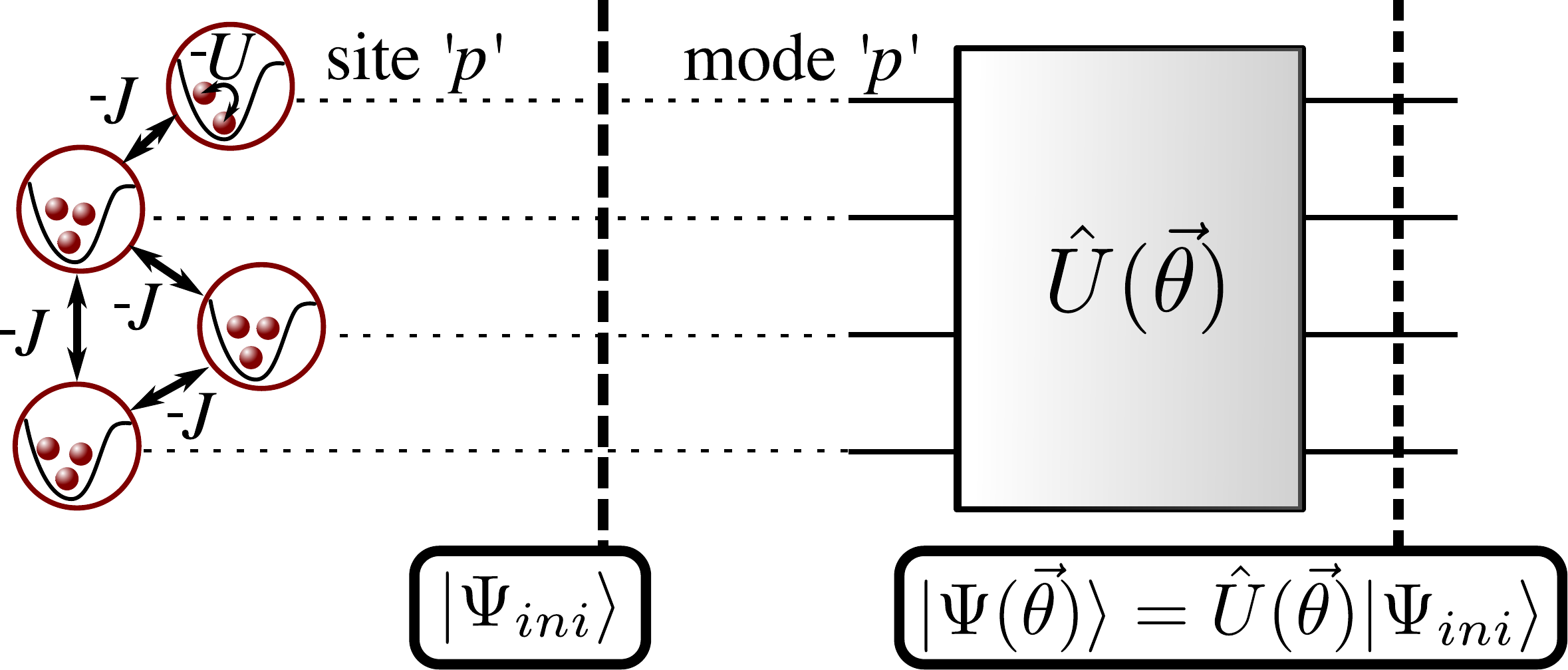}
    \caption{\textbf{Illustration of the VQA approach.} Each photonic mode in the quantum circuit is used to encode the bosonic occupation of a particular site in the Bose-Hubbard network.}
    \label{fig:site_to_mode}
\end{figure}

In contrast to most of the studies focusing on qubit devices, we tackle the problem of many-bosons systems using CV circuits.
The choice of CV quantum computers is rather intuitive as there exists a straightforward one-to-one correspondence between the modes of a bosonic system and the photonic modes of such devices. 
In the case of the attractive BH system, each photonic mode $p$ of the quantum circuit can be used to encode the bosonic population of the associated site $p$, as illustrated in Fig.~\ref{fig:site_to_mode}.
Starting from this advantageous property, we will introduce two quantum ansatze based on CV photonic gates to encode many-boson wavefunctions for VQAs.
 
\subsubsection{Utilized CV quantum gates}

 \begin{figure}[h]
    \centering
    \includegraphics[width=\columnwidth]{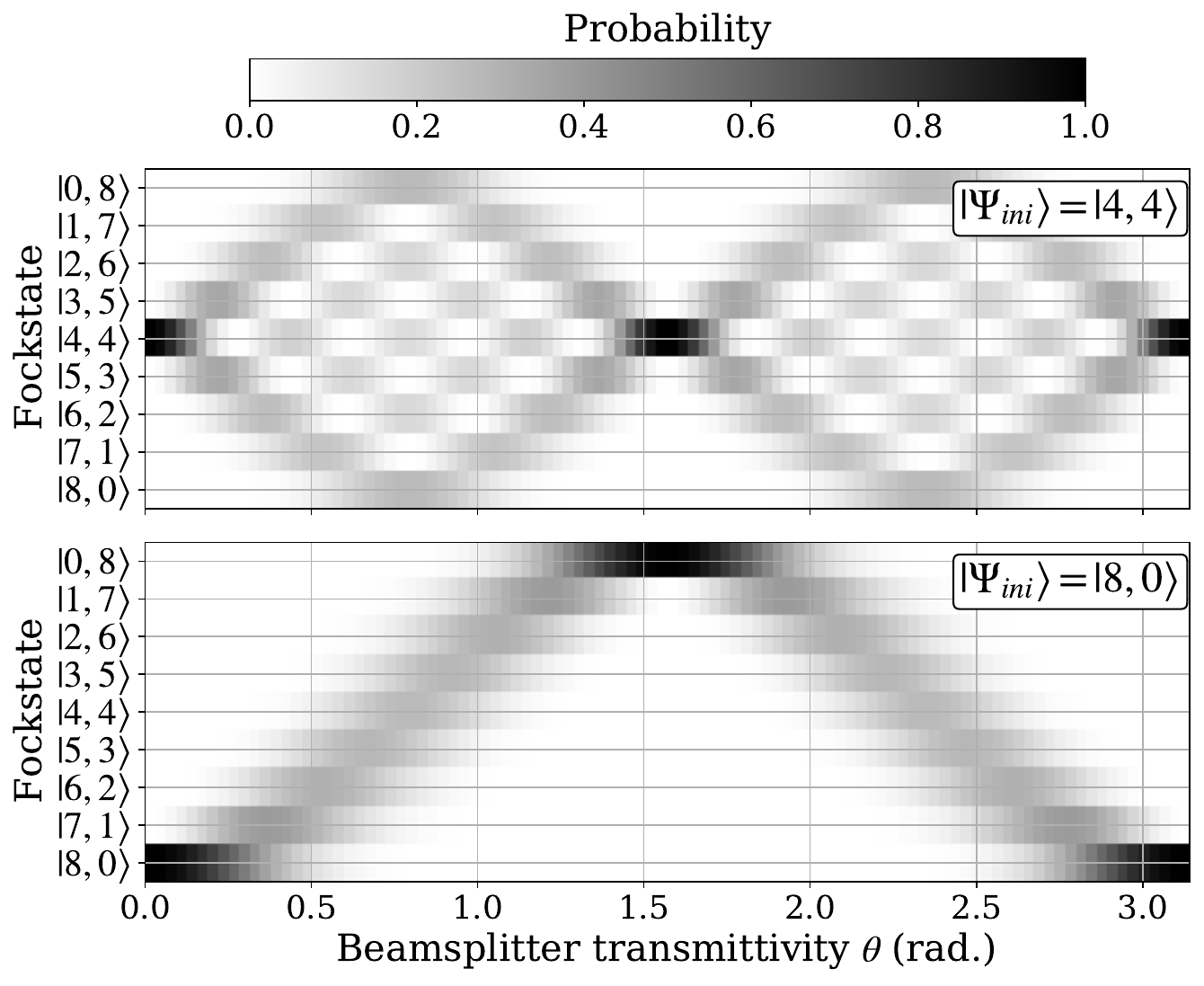}
    \caption{ \textbf{Illustration of the effects of a BS-gate.} Density of probability of a state $B(\theta,0)\ket{\Psi_{ini}}$ (with $\ket{\Psi_{ini}}$ an initial state) in the Fock basis as a function of the transmittivity parameter $\theta$ of the beam-splitter unitary. The probability is illustrated with shades of grey, ranging from white for zero to black for unity. \textbf{Upper panel:} $\ket{\Psi_{ini}}=\ket{4,4}$. \textbf{Lower panel:} $\ket{\Psi_{ini}}=\ket{8,0}$. }
    \label{fig:BS_spreading}
\end{figure}
 
Concerning the fundamental properties of both ansatze, we know that the BH Hamiltonian in Eq.~\eqref{eq:HBH} is boson-number-preserving by construction (i.e.~$N_B$ is constant).
Thus, in order to preserve this property, we choose for both ansatze to consider only three types of photon-number-preserving gates, namely: beam-splitters, rotation gates and Kerr gates.
First, the beam-splitter gate is a two-mode gate 
that affects two inputs modes simultaneously (labeled by $p$ and $q$)
with an unitary defined as
\begin{align}
    B_{p,q}(\theta,\phi) = \exp \left( \theta(e^{i\phi}a_q^\dagger a_p- e^{-i\phi} a_p^\dagger a_q )\right),
\end{align}
where $\theta$ and $\phi$ are respectively the transmittivity and phase angle of the gate. The action of the beam-splitter's unitary on the associated ladder operators ($a_p/a_p^\dagger$ and $a_q/a_q^\dagger$) is given by
\begin{equation}\label{eq:BS_action}
\begin{split}
    B_{p,q}(\theta,\phi)^\dagger a_p B_{p,q}(\theta,\phi) &= a_p \cos(\theta) - a_q \sin(\theta) e^{-i\phi},\\
    B_{p,q}(\theta,\phi)^\dagger a_q B_{p,q}(\theta,\phi) &= a_q \cos(\theta) + a_p \sin(\theta) e^{i\phi}.
\end{split}
\end{equation}
To have a better illustration of the effects of the beam-splitter gate on a many-photon wavefunction, figure~\ref{fig:BS_spreading} shows the evolution of the density of probability of a two-mode state $B(\theta,0)\ket{\Psi_{ini}}$ (with $\ket{\Psi_{ini}}=\ket{8,0}$ or $\ket{4,4}$) in the Fock basis as a function of the transmittivity parameter $\theta$. 
As shown in this figure, the main effect of the beam-splitter is to delocalize the photons from a mode to another one creating various patterns in the Fock basis. This strongly depends on the initial state considered. The phase rotation gate acts on a Fock state as $R_p(\theta) \ket{n_p} =  e^{i\theta n_p }  \ket{n_p}$ with the unitary operator  
\begin{align}
       R_p(\theta) = \exp\left( i\theta a_p^\dagger a_p \right).
\end{align} 
Finally, the Kerr gate is a non-Gaussian single-mode gate whose unitary reads
\begin{align}
       K_p(\theta) = \exp\left( i\theta (a_p^\dagger a_p)^2 \right).
\end{align}
For a given modes $p$, this leads to $K_p(\theta) \ket{n_p} =  e^{i\theta n_p^2 }  \ket{n_p}$,
thus generating a non-linear phase term that depends on the squared number of bosons $n_p^2=(a_p^\dagger a_p)^2$ present in the mode $p$. 
Although the implementation of Kerr gates as an optical component is a true challenge on current devices, the incessant development of new techniques and non-linear materials tends to suggest that reliable ways of implementing these gates will be available in the near future~\cite{azuma2007quantum, sefi2013measurement,hillmann2020universal}.
For the purpose of our theoretical work, we assumed the possibility to implement these gates without any practical constraints. This assumption was motivated by our analysis which revealed the key role played by the Kerr interaction in the performance of our ansatze (as it allows us to go beyond Gaussian transformations).

 Note that the choice of the three types of gates presented here is not only motivated by their capacity to preserve a same number of bosons $N_B$, but also by their physical action on a many-boson wavefunctions. Indeed, the role of a BS gate is essentially to explore the space of many-boson configurations for two connected modes.  On another note, the addition of Kerr (or rotation) gates in a circuit after a BS gate makes it possible to non-linearly affect the phase of each many-bosons configurations of the wavefunction (see Appendix~\ref{appendix:illustration} for an illustration of this with the minimal ansatz). These behaviors were the major motivations for the design of the ansatze introduced in the next section, that can therefore be regarded as physically-motivated as well as hardware-efficient.

\subsubsection{Minimal beam-splitter-Kerr ansatz}
 
\begin{figure}
    \centering 
    \includegraphics[width=4cm]{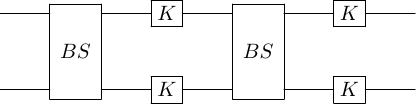}\\ \vspace{0.5cm}
    \includegraphics[width=6cm]{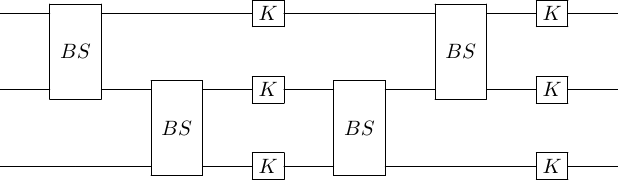}\\ \vspace{0.5cm}
    \includegraphics[width=\columnwidth]{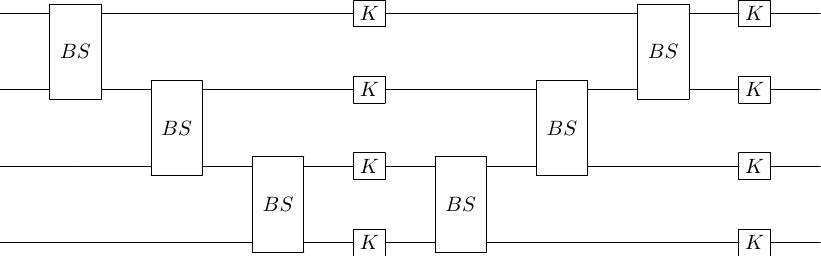}
    \caption{\textbf{BS-Kerr ansatz.} Quantum circuit structure for $N_S=2$, 3 and 4 sites with two consecutive layers. Each layer is composed of a series of $N_S-1$ beam-splitter gates in stairs (up or down), followed by a series of $N_S$ Kerr gates applied to each modes. The total gate count is $N_L(2N_S-1)$ where $N_L$ is the number of layers.}
    \label{fig:minimalcircuit}
\end{figure}

Let us now introduce the first ansatz developed to encode the ground state of the attractive BH model: the so-called {\it minimal beam-splitter-Kerr} (BS-Kerr) ansatz.
This circuit architecture was designed to produce a high fidelity encoding with as few quantum gates and parameters to optimize as possible, based on numerical investigations.
Considering these constraints, the most ``efficient'' architecture we found (in terms of the total number of gates required to encode the ground state with sufficient accuracy) resulted in layers of beam-splitter and Kerr gates as shown in Fig.~\ref{fig:minimalcircuit} for
two consecutive layers.
Note that the phase angle parameters of the beam-splitters are always set to  $\phi=0$ in this ansatz. 
Indeed, we did not see any improvement
when considering the latter in our numerical simulations (not shown).
Discarding these angles allows to reduce the complexity of the classical optimization with no impact on the final accuracy.

As readily seen in Fig.~\ref{fig:minimalcircuit},
each layer is defined as a series of $N_S-1$ beam-splitter gates in stairs (up or down) followed by a series of $N_S$ Kerr gates applied on each mode, thus leading to a gate count of $N_L(2N_S-1)$
(and the same number of parameters to optimize)
where $N_L$ is the number of layers.
From a layer to another one, we alternate the staircase structure to ensure that each mode can have the opportunity to exchange photons (via the beam-splitters) with all the other modes in the circuit. 


Note that removing the Kerr gates from the ansatz leads to very poor results (not shown), showing that a passive Gaussian transformation is insufficient to encode the ground state of the attractive BH model despite we considered non-Gaussian inputs.

\begin{figure}[t]
    \centering 
    \includegraphics[width=5cm]{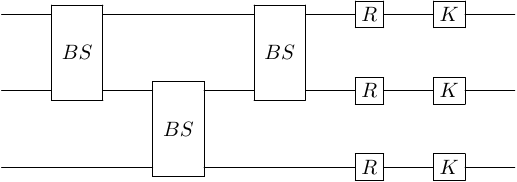}\\ \vspace{0.5cm}
    \includegraphics[width=6cm]{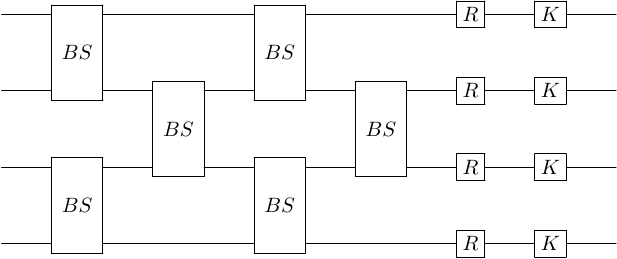}
    \caption{\textbf{Interferometer-Kerr ansatz.} Quantum circuit architecture (a single layer)  for $N_S = 3$ (top) and $N_S = 4$ (bottom) modes.}
    \label{fig:interferometerKerr}
\end{figure}

\subsubsection{Interferometer-Kerr ansatz}\label{sec:intK_ansatz}

The BS-Kerr ansatz was designed to encode the ground state of the attractive BH model with a minimal number of gates and parameters, which is advantageous for numerical simulations.
However, in experimental realizations one may want to minimize the number of Kerr layers. Therefore, we considered an alternative ansatz composed of layers of one interferometer and one Kerr gate per mode, as depicted in Fig.~\ref{fig:interferometerKerr} for 3 and 4 modes.
The interferometer is composed of $N_S(N_S - 1)/2$ beam-splitters and $N_S$ rotation gates, making a total of $N_L N_S(N_S + 3)/2$ gates and $N_L N_S(N_S + 1)$ parameters (as one beam-splitter accounts for two parameters).

At first sight, this ansatz appears less efficient than the BS-Kerr ansatz in terms of number of gates and parameters per layer.
However, we can expect the additional expressibility brought about by the complete interferometers to lead to a reduction in the number of layers required to encode the ground state.
In other words, the interferometer-Kerr ansatz could lead to a reduction in the total number of Kerr gates at the expense of additional beam-splitter and rotation gates, which are however much simpler and cheaper.
This hypothesis is numerically investigated in Sec.~\ref{subsec:interferometer_res}.

The complexity of the ansatz can be reduced by getting rid of the rotation gates (physically irrelevant for most applications~\cite{clements2016optimal}) and/or setting the phase angles of all beam-splitters to $\phi=0$ (and keeping only the transmittivity angles) as already done for the BS-Kerr ansatz.

\begin{figure*}
    \centering
    \includegraphics[width=\textwidth]{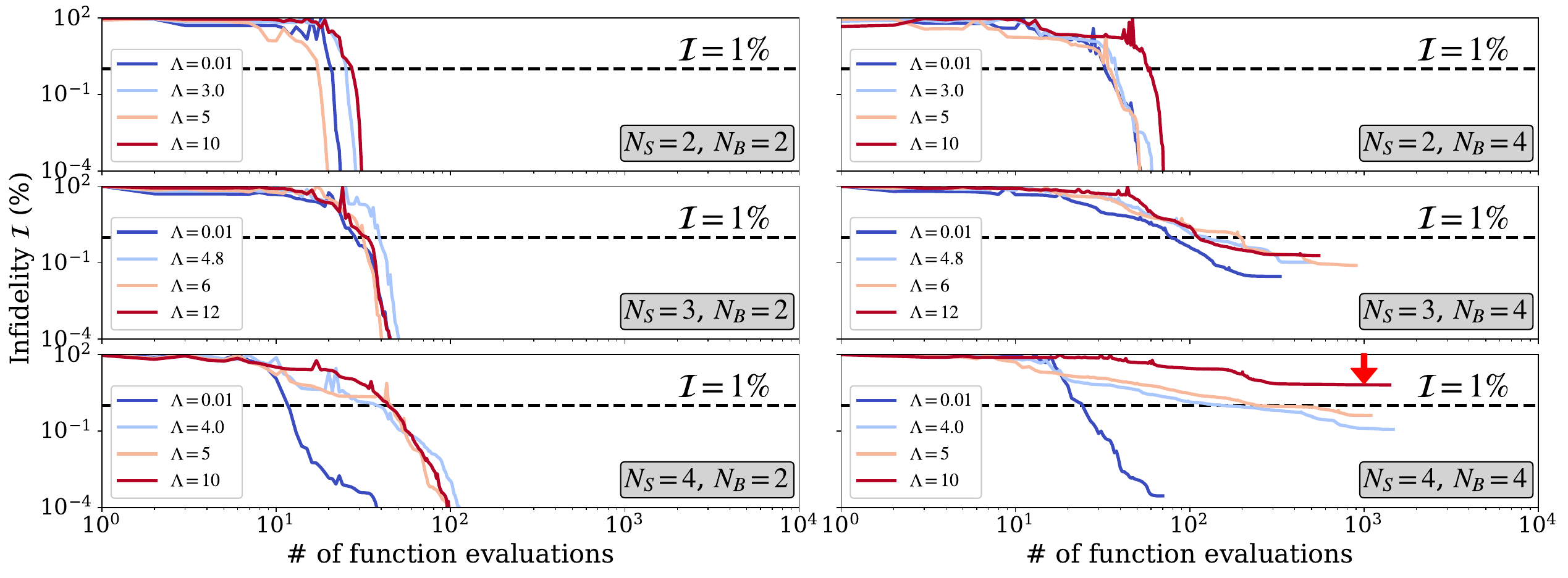}  
    \caption{
    {
    \textbf{Illustration of infidelity optimization runs realized to study the expressibility of the BS-kerr ansatz.} Each plot shows the evolution of the infidelity $\mathcal{I}$ as a function of the number of function evaluations. These results were obtained for $N_L=5$ layers circuits and initial single-mode states (i.e.~$\ket{\Psi_{ini}} = \ket{N_B,0}$, $\ket{N_B,0,0}$ and $\ket{N_B,0,0,0}$ for the dimer, the three-, and the four-site systems, respectively). Left plots show results for $N_B=2$, whereas right plots for $N_B=4$. Each row (from top to bottom) shows results for the dimer, the three-, and the four-site BH models, respectively. In each case, four values of $\Lambda$ have been considered to span the different regimes of many-body correlations (as marked in figures~\ref{fig:dimer_GS} and \ref{fig:3_4sites}).
    The infidelity limit $\mathcal{I} = 1\%$ considered as an indicator of the good ansatz expressibility is marked by a horizontal black dashed line. Note that all the cases treated here reach this limit except one  which is stressed by a red arrow (on the bottom right panel). 
    }
    } \label{fig:optimization_illustration}
\end{figure*}

\section{Numerical investigation of the ansatze capabilities}\label{sec:num_results_expressibility}

 In this section, we present a numerical investigation of the expressibility (and resource efficiency)  for the two ansatze introduced previously applied to the BH dimer, the three-site and the four-site periodic chains.
The simulations of the photonic circuits were realized using the Python package Strawberryfields~\cite{killoran2019strawberry}. 
For the optimization of the quantum circuits parameters, the L-BFGS-B method was used as encoded in the optimParallel package~\cite{optimpar}.
The circuit's parameters were always initialized with a small random amplitude drawn in $[-0.05,+0.05]$, and the maximum number of iterations was set to 20000. An ED code has been implemented in Python to compute the exact ground state by exact diagonalization. Further we also provide a sample code with the implementations of all the components of our method in \cite{code_BHCV} for interested readers.

\subsection{Capacity of the BS-Kerr ansatz}\label{subsec:minimal_res}

To demonstrate and characterize the capacity of the different ansatze to encode the ground state of BH networks, we estimated the fidelity of the trial state with respect to the exact ground state of the system $\ket{\Psi_0}$,
\begin{equation}
    \mathcal{F} = |\braket{\Psi_0 |\hat{U}(\vec{\theta}) |\Psi_{ini}}|^2.
    \label{eq:fidelity}
\end{equation}
Assuming that the classical optimization process does not end up in local minima,
an ansatz with a strong encoding capability will return a fidelity $\mathcal{F}$ close to unity. 
We numerically check this property by optimizing the quantum circuit's parameters to minimize a complementary measure that is the infidelity
\begin{eqnarray}
\mathcal{I} = 1-\mathcal{F}
\label{eq:infidelity}
\end{eqnarray}
which is naturally lower-bounded by zero. 
Note that no mention has been made so far about the nature of the initial state $\ket{\Psi_{ini}}$, for which we considered two choices: a single-mode and a two-mode states, detailed later in section~\ref{sec:BHdimer}.

{
\subsubsection{Illustration of the infidelity optimisation}\label{sec:illustration_optimisation}
}

{
As an introduction of our study, in this first subsection we illustrate the evolution of the variational minimization of the infidelity by the mean of representative single runs of optimization. 
We gather in Fig.~\ref{fig:optimization_illustration} several examples of optimization runs showing the minimization of the measure $\mathcal{I}$ as a function of the number of function evaluations called in the L-BFGS-B algorithm (as provided by the optimParallel package~\cite{optimpar}), for different values of $\Lambda$ to span the different correlation regimes.
Each row of Fig.~\ref{fig:optimization_illustration} (from top to bottom) shows results for the dimer, the three and four-site BH model respectively, for $N_B=2$ (left panels) and $N_B=4$ (right panels) bosons.
$N_L = 5$ layers of the BS-Kerr ansatze were considered for all optimization runs, whatever the size of the system. As a remark, note that the optimization runs presented here are essentially illustrative.
The shape of the optimization curves may actually differ from a run to another as small initial random values of circuit parameters are considered in practice.
We checked the effects of these fluctuations
by realizing multiple optimizations with random initial parameters and we noticed negligible changes in the final results obtained, hence the tendencies shown in Fig.~\ref{fig:optimization_illustration} can be regarded as sufficiently general.
}

{As readily seen in Fig.~\ref{fig:optimization_illustration}, every optimization curve crosses the limit $\mathcal{I} = 1 \%$ (defined as the marker of good expressibility for the ansatze)
for all values of $\Lambda$ but $\Lambda = 10$ on the last panel.
The left panels ($N_B=2$) manifest very low converged infidelities,
and thus a very accurate encoding of the ground state of the system. 
The number of function evaluations required to fulfill the condition $\mathcal{I} = 1 \%$ (and to reach final convergence) ranges from around $10$ to $100$ in these cases.
Conversely, increasing the number of bosons in the systems leads to an increase of the number of function evaluations, as illustrated on the right panels ($N_B=4$).
In this case, the number of evaluations required to satisfy the condition $\mathcal{I} = 1 \%$ ranges from around $20$ to $300$. 
To reach final convergence (i.e. when the selected optimizer cannot minimize the cost function anymore) the number of evaluations also increases, ranging from $50$ to $2000$ approximately (all systems considered).
The number of function evaluations increases
with the number of bosons $N_B$ and sites $N_S$,
i.e. with the size of the Fock space $\mathcal{D}_{\hat{\mathcal{H}}}$
(as shown in table~\ref{table:dimensions}) that is related to the complexity of the encoding problem. 
As a remark, note that in practice the final number of function evaluations $N_{feval}^{fin}$ differs from the final number of iterations $N_{iter}^{fin}$ realized by the L-BFGS-B algorithm, as one iteration of the algorithm requires multiple function evaluations. 
We choose an upper limit of 20000 iterations to ensure convergence of the algorithm whatever the case considered.
And as illustrated in Fig.~\ref{fig:optimization_illustration}, this limit is never reached as $N_{iter}^{fin} \leq N_{feval}^{fin} \ll 20000$ (a feature that we also checked in many other test simulations).
In Fig.~\ref{fig:optimization_illustration}, we see that a circuit with a depth of $N_L=5$ layers of the BS-Kerr ansatz is sufficient to ensure an accurate encoding of all examples considered, except
for the strongly interacting case of the largest system $N_S=4$, $N_B=4$ and $\Lambda = 10$.
In this case, the converged infidelity is $\mathcal{I} \approx 7 \%$ (i.e.~$\mathcal{F} \approx 93\%$).
A much better fidelity can be reached by adding more layers to the quantum circuit, thus enhancing the expressibility of the ansatz. 
Typically, changing $N_L$ from 5 to 8 (not shown here), makes all the examples below the limit $\mathcal{I}=1\%$.}

{In spite of these promising results, one can still suspect the presence of barren plateaus that appear for large systems in VQAs when random initial parameters are used.
For qubit-based devices, barren plateaus denote the exponentially vanishing cost function gradients with the number of qubits~\cite{mcclean2018barren,sharma2020trainability,cerezo2021cost,arrasmith2020effect,uvarov2021barren,cerezo2021higher}.
Note that the exact conditions for the appearance of barren plateaus on qubit-devices
remains unclear, as several recent works have shown that they can
also arise when physically-inspired~\cite{larocca2021diagnosing,wang2020noise} and beyond 2-design ansatze~\cite{holmes2021connecting} are considered.
It is an open question of how and when barren plateaus may appear in CV ansatze, although recent studies indicate that CV systems may exhibit barren plateaus where the cost gradients vanish
exponentially with the number of system modes (instead of the number of qubits)~\cite{volkoff2020efficient,volkoff2021universal}.
In this work, we did not see any optimization problem that could indicate the presence of such phenomenon,
but it remains unclear for larger systems.
Fortunately, due to the analogy between the barren plateaus phenomenon on discrete and continuous variable systems~\cite{volkoff2020efficient},
one can expect that strategies used to bypass the barren plateaus on qubit-based devices~\cite{haug2021optimal,patti2020entanglement,larocca2021diagnosing} are also transferable to CV systems.
Such an involved study is out of the scope of the manuscript and is left for future work.}

{
In the next sections, we will present the results of our numerical study of the expressibility (and the resource efficiency) of the ansatze as a function of the number of bosons $N_B$ and sites $N_S$. 
Our approach was inspired by Ref.~\cite{cade2020strategies}.
Namely, we realize a series of simulations where we progressively increase the number of layers $N_L$ in the quantum circuits and perform each time one optimization run of infidelity. The choice of single runs was motivated by many numerical simulations which shown that very similar results are always obtained in terms of encoding capacity and resource efficiency. 
In practice, the more the number of layers $N_L$ increases, the more the expressibility of the ansatz improves.
We continue to increase the number of layers $N_L$ until the infidelity of the ground state fulfill the condition $\mathcal{I} \leq 1\%$ (i.e.~$\mathcal{F} \geq 99\%$).
}

\subsubsection{BH model with $N_S=2$ sites}\label{sec:BHdimer}

\begin{figure}[t]
    \centering
    \includegraphics[width=\columnwidth]{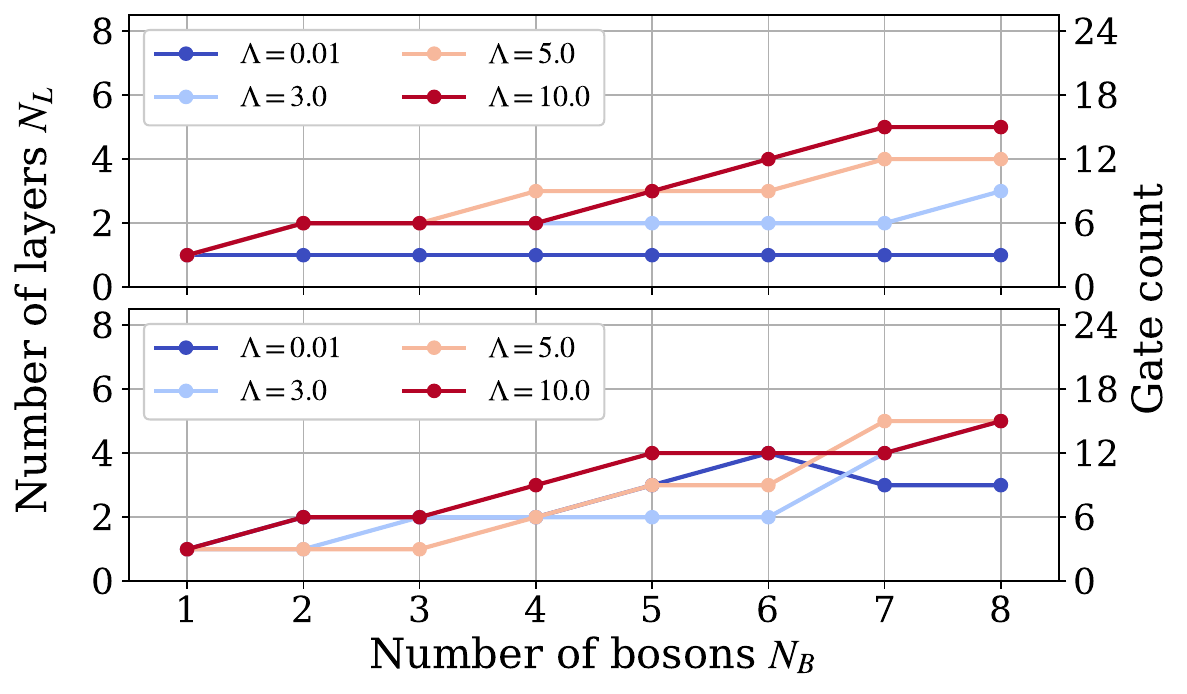}
    \caption{\textbf{Ansatz encoding capacity (dimer)}. Minimum depth required (in terms of number of layers $N_L$ and gate count) to generate a fidelity $\mathcal{F} \geq 99\%$ [see Eq.~(\ref{eq:fidelity})] for different values of $\Lambda$ and $N_B$. \textbf{Upper panel:} results obtained with the single-mode initial state. \textbf{Lower panel:} results obtained with the two-mode initial state. }
    \label{fig:Proof_dimer}
\end{figure}

Focusing first on the BS-Kerr ansatz for the BH dimer, the minimum number of layers detected to reach $\mathcal{F} \geq 99\%$ for $N_B=1$ to $N_B=8$ is shown in Fig.~\ref{fig:Proof_dimer}
for $\Lambda$ values corresponding to different regimes of correlation (as marked by the vertical red-dashed lines in Fig.~\ref{fig:dimer_GS}).
Two different initial states were considered: a single-mode one given by $\ket{\Psi_{ini}} = \ket{N_B,0}$ and a two-mode one given by $\ket{\Psi_{ini}} = \ket{\frac{N_B}{2},\frac{N_B}{2}}$ (\textit{n.b.} for odd total number of bosons we took $\ket{\Psi_{ini}} =\ket{\frac{N_B+1}{2},\frac{N_B+1}{2}-1}$ instead). 
{As a first general remark, Fig.~\ref{fig:Proof_dimer} shows that, with only a few layers of circuit, a fidelity higher than 99$\%$ is always reached whatever the case considered (i.e.~the correlation strength $\Lambda$, the number of bosons $N_B$ and the choice of initial state).
These numerical results give a first demonstration that short depth CV circuits can accurately encode many-boson wavefunctions even in the strongly correlated regime, at least in the case of system sizes we have considered.}
Note that to ensure an accurate encoding of the wavefunction, the larger $N_B$ is, the larger the number of layers $N_L$ needs to be.
This is expected as the complexity of the ground state grows with the size of the Fock space $D_{\hat{\mathcal{H}}}$, which is linear in $N_B$ for the BH dimer (as shown in Tab.~\ref{table:dimensions}).
Interestingly, the trend of the curves ensuring an efficient encoding is globally sub-linear in $N_B$ whatever the initial states considered. Roughly, we see here that choosing a number of layers such as $N_L \approx 0.7\times N_B$ seems to guarantee a very high fidelity in all cases considered herein. {The maximum number of Kerr gates involved here reaches a maximum of $N_{Kerr} = 10$ (as this number scales like $N_{Kerr} = N_SN_L$).
}

On another note, an interesting fact arises in the weakly correlated case $\Lambda = 0.01$ when the single-mode initial state is used (dark blue curve in the upper panel). In this case, a single layer of the BS-Kerr ansatz is sufficient to reach a fidelity higher than 99$\%$, even when the number of bosons increases.
This property is true only for the single-mode initial state.
We rationalize this result in Appendix~\ref{app:demo}, where
we show that, starting from such a state, an exact encoding ($\mathcal{F} = 100\%$) of the BH ground state {in the non-interacting (superfluid) regime ($\Lambda \rightarrow 0$)} can be realized using only beam-splitters (this demonstration is realized for the three networks considered in our study).
{Note that even if no Kerr gates are required in this particular limit, an exact classical simulation of Fock states propagating in a passive interferometer has exponential complexity, as shown by seminal papers on the computational complexity of linear optical quantum computing~\cite{aaronson2013computational,gard2014inefficiency}.
Indeed, even if the interferometer itself can be fully described by a 2$N$-dimensional orthogonal matrix, (which is tractable), the action of an interferometer on an arbitrary input state requires a Fock space description of the gate (as opposed to a Gaussian input state, not considered here).}

\begin{figure}[t]
    \centering
    \includegraphics[width=\columnwidth]{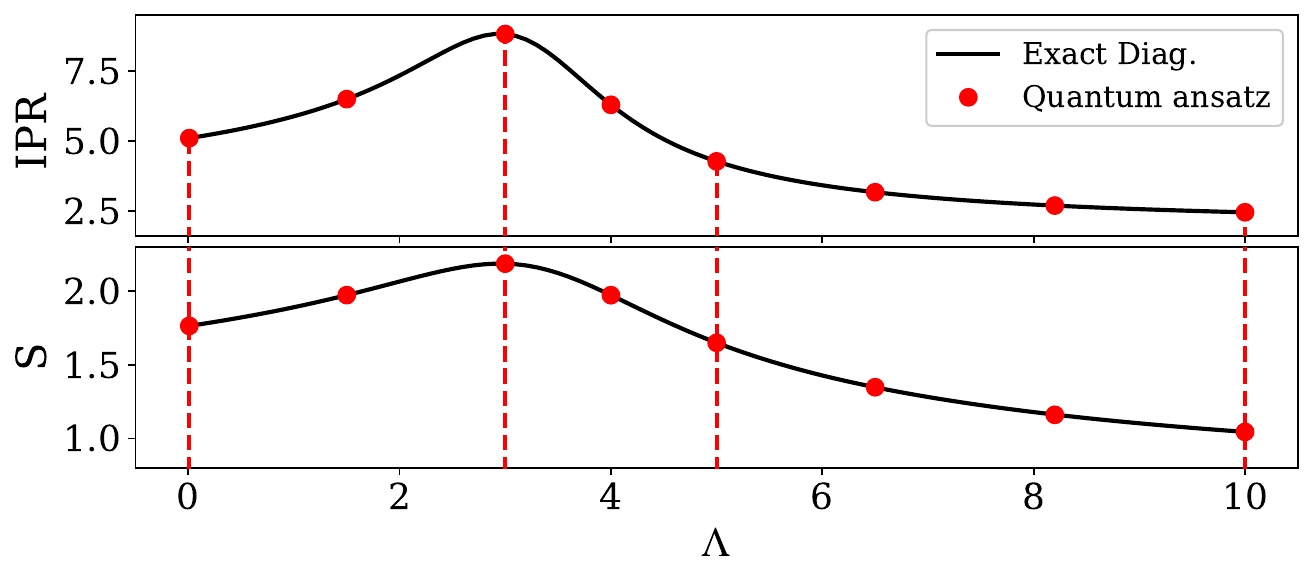}
    \includegraphics[width=\columnwidth]{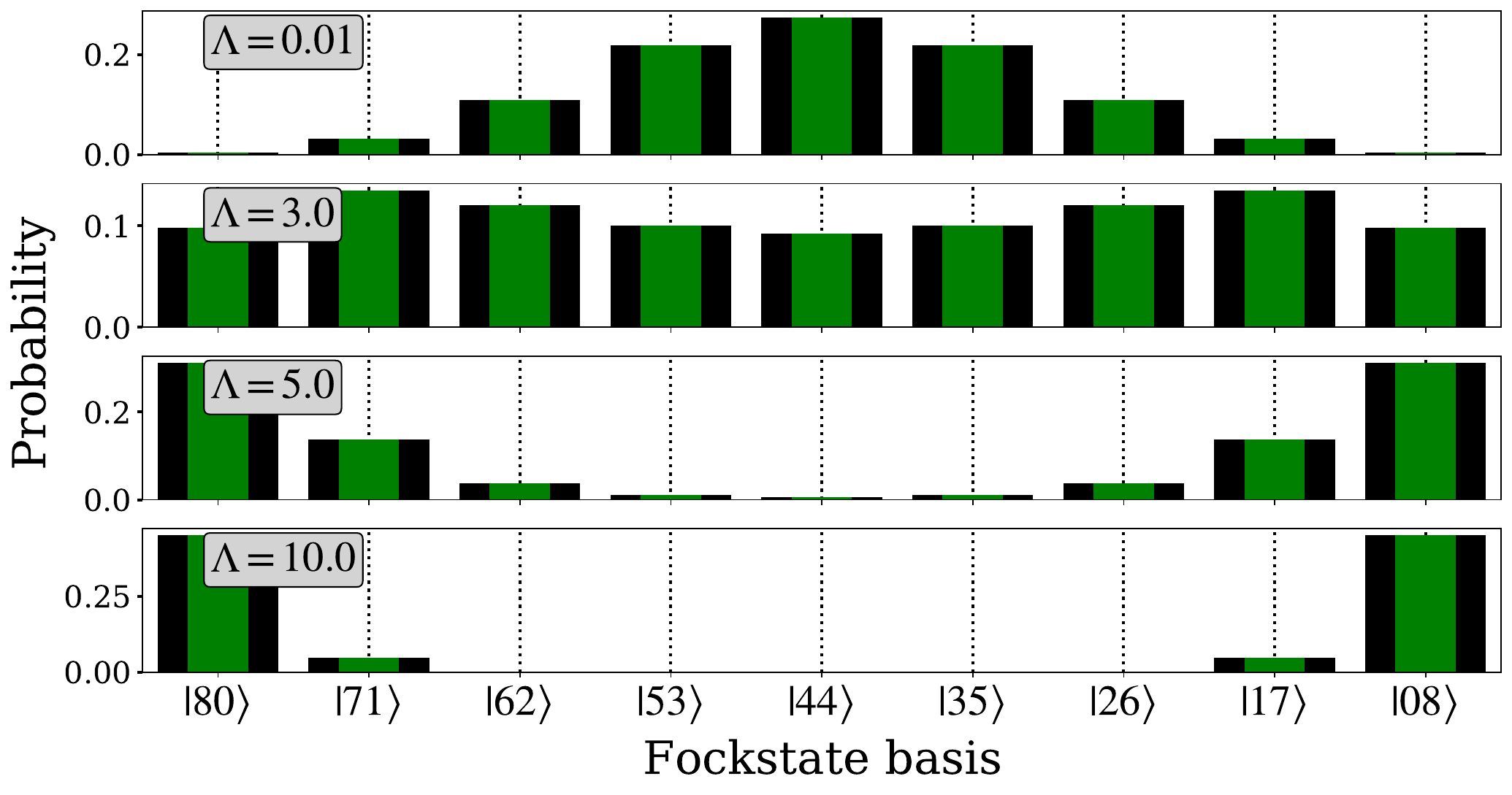}
    \caption{
    {
    \textbf{Trial-state properties (dimer)}. Illustration of the properties of the quantum state obtained by minimization of the infidelity, using the BS-Kerr ansatz with $N_L=6$, $N_B=8$, $\ket{\Psi_{ini}} = \ket{4,4}$. \textbf{Upper panel:} IPR and entanglement entropy $S$ of the resulting ansatz state (red dots) compared to the exact ground state (black curves) as a function of $\Lambda$.
    \textbf{Lower panel: } Probability of occupation of the ansatz state in the Fock basis (green bars) compared with exact diagonalization (black bars). The $\Lambda$ values chosen for the four configurations are marked in the upper panel with red dashed vertical lines (same as in Fig.~\ref{fig:dimer_GS}).
    }
    }
    \label{fig:dimer_fidelity}
\end{figure}

As an addition, in Fig.~\ref{fig:dimer_fidelity} we illustrate the evolution of the structural properties of a trial state (built with the BS--Kerr ansatz) as a function of $\Lambda$. In this case, we focus on a system with $N_B = 8$ bosons and we use a quantum circuit of $N_L=6$ layers with $\ket{\Psi_{ini}} = \ket{4,4}$ as an initial state. The results obtained are compared to the exact ones generated via ED.  On the upper panel of Fig.~\ref{fig:dimer_fidelity}, we show the evolution of the IPR and the entanglement entropy $S$ of the trial state (red dots) compared to exact results (black curve) as a function of $\Lambda$. Note that each of the trial states obtained in the simulations have a fidelity $\mathcal{F} \geq 99 \%$. The lower panel shows the evolution of the probability of occupation in the Fock basis for the trial state (green bars) compared to the exact ground state (black bars) for four representative values of interaction parameter $\Lambda=0.01, 3, 5$ and $10$.  The results presented here highlight the structural transformation of the dimer ground state as $\Lambda$ increases, starting from the superfluid regime ($\Lambda = 0.01$) and going through the maximum spreading structure ($\Lambda = 3$ for a maximal IPR value) to finally reach the famous ``cat-like'' superposition (with $\Lambda =5$ to 10 and the IPR tending to $2$).
For all values of $\Lambda$ considered, the IPR, the entanglement entropy $S$ and the probability of occupation of the trial states are in very good agreement with the exact results, thus showing that the expressibility of our ansatz is strong enough to reproduce the exotic features of the exact ground state in different correlation regimes. Similar illustrations of the BH ground state restructuring will be given in the following of the paper for the three and four-site BH system.

\subsubsection{BH model with $N_S=3$ and $4$ sites}

 \begin{figure}[t]
    \centering 
    \includegraphics[width=\columnwidth]{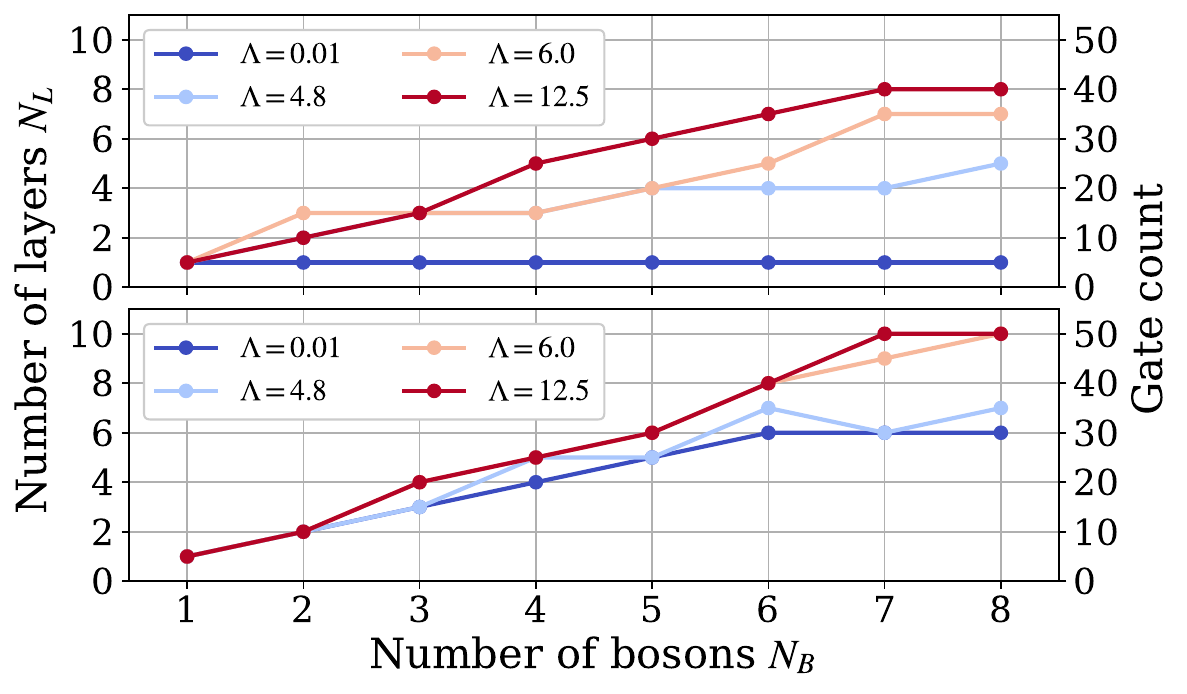}
\caption{\textbf{Ansatz encoding capacity (3-site chain).} Minimum depth required (in terms of number of layers $N_L$ and gate count) to generate a fidelity $\mathcal{F} \geq 99\%$ [see Eq.~(\ref{eq:fidelity})] for different values of $\Lambda$ and $N_B$. \textbf{Upper panel:} results obtained with a single-mode initial state. \textbf{Lower panel:} results obtained with a two-mode initial state.}
    \label{fig:Proof_3sites}
\end{figure}


Turning to larger systems, Figs.~\ref{fig:Proof_3sites}  show the results obtained on the three-site BH model
for $\Lambda$ values corresponding to different regimes of correlation (as marked by the vertical red-dashed lines in the left panel Fig.~\ref{fig:3_4sites}).
A single-mode and two-mode initial states are again considered and read $\ket{\Psi_{ini}} = \ket{N_B,0,0}$ and $\ket{\Psi_{ini}} = \ket{\frac{N_B}{2},0,\frac{N_B}{2}}$ respectively (\textit{n.b.} for odd total number of bosons we took $\ket{\Psi_{ini}} = \ket{\frac{N_B+1}{2},0,\frac{N_B+1}{2}-1}$ instead).
Compared to the BH dimer, the dimension of the Fock space is non-linear in $N_B$ (see Tab.~\ref{table:dimensions}) thus leading to a maximum number of bosonic configurations accessible of $D_{\hat{\mathcal{H}}}=45$ when $N_B=8$ (approximately 6 times larger than the number of configurations on the dimer for the same number of bosons). 
In spite of this, we observed that the BS-Kerr ansatz performs very well as shown in Fig.~\ref{fig:Proof_3sites}. 
In this case, we retrieve some differences between the results obtained with the two initial states. 
The single-mode initial state seems to perform better in term of resource efficiency as the latter requires globally less layers to encode the ground state with a fidelity higher than 99$\%$. {Typically, we see in Fig.~\ref{fig:Proof_3sites} that the number of layers detected increases up to $N_L=8$ for a single-mode initial state against a maximum of $N_L=10$ layers for the two-mode one. The associated number of Kerr gate being 24 and 30 respectively.} Still, the results obtained here reveal the good expressibility of the minimal ansatz which is able to accurately transform both different initial states into very good approximations of the exact ground state. 
As readily seen in Fig.~\ref{fig:Proof_3sites}, choosing a circuit depth with $N_L \sim 1.25 \times N_B$ guaranties an efficient encoding of the ground state in all cases considered here. As a last remark for the three-site system, note again that only one layer is required for $\Lambda = 0.01$ when this initial state is used (dark blue curve). We refer the interested reader to Appendix.~\ref{app:demo} where we give a mathematical demonstration of this feature. 

 \begin{figure}[t]
    \centering 
\includegraphics[width=\columnwidth]{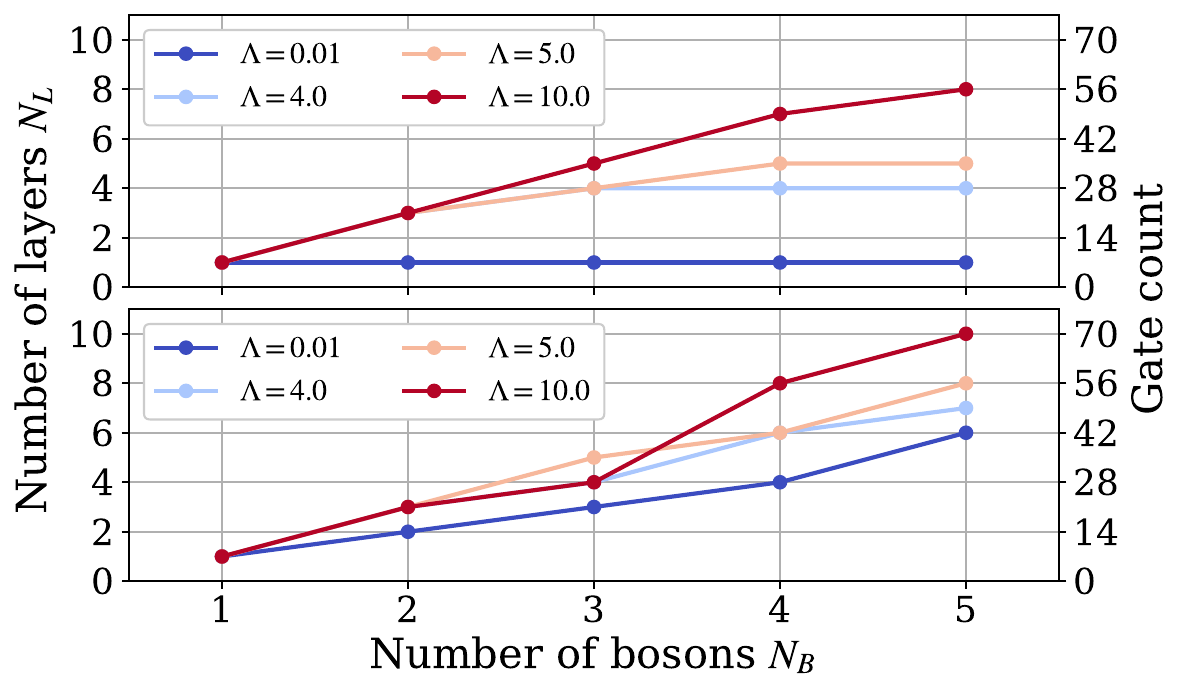}
\caption{\textbf{Ansatz encoding capacity (4-site chain).} Minimum depth required (in terms of number of layers $N_L$ and gate count) to generate a fidelity $\mathcal{F} \geq 99\%$ [see Eq.~(\ref{eq:fidelity})] for different values of $\Lambda$ and $N_B$. \textbf{Upper panel:} results obtained with a single-mode initial state. \textbf{Lower panel:} results obtained with a two-mode initial state.}
    \label{fig:Proof_4sites}
\end{figure}

  \begin{figure}[t]
    \centering
    \includegraphics[width=\columnwidth]{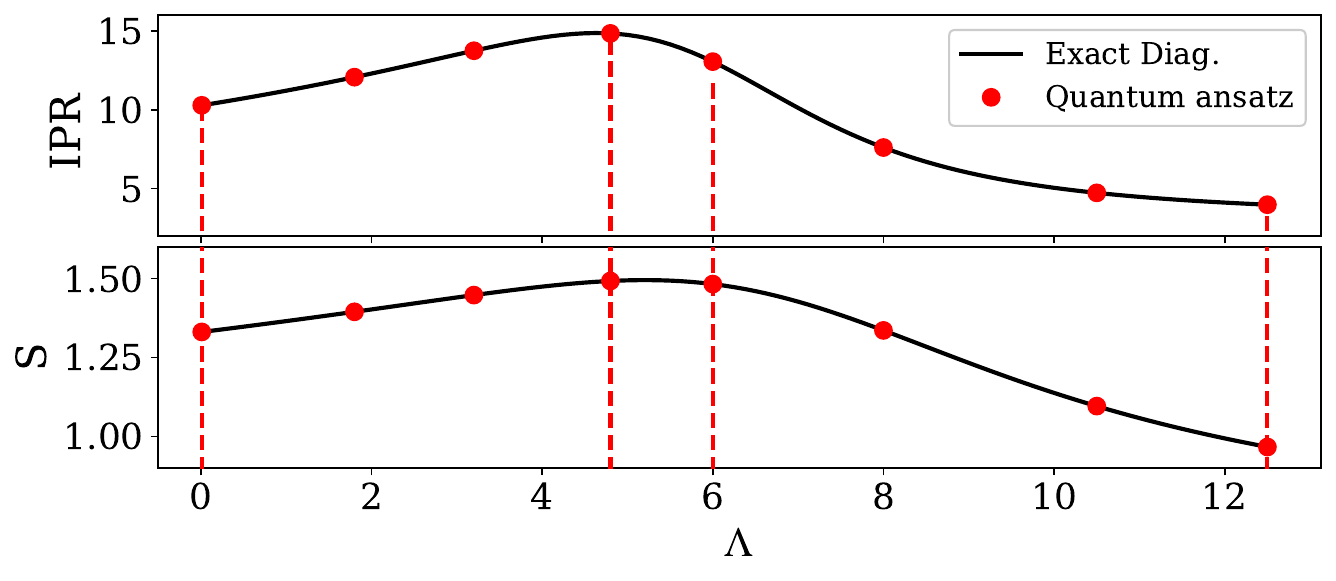}
    \includegraphics[width=\columnwidth]{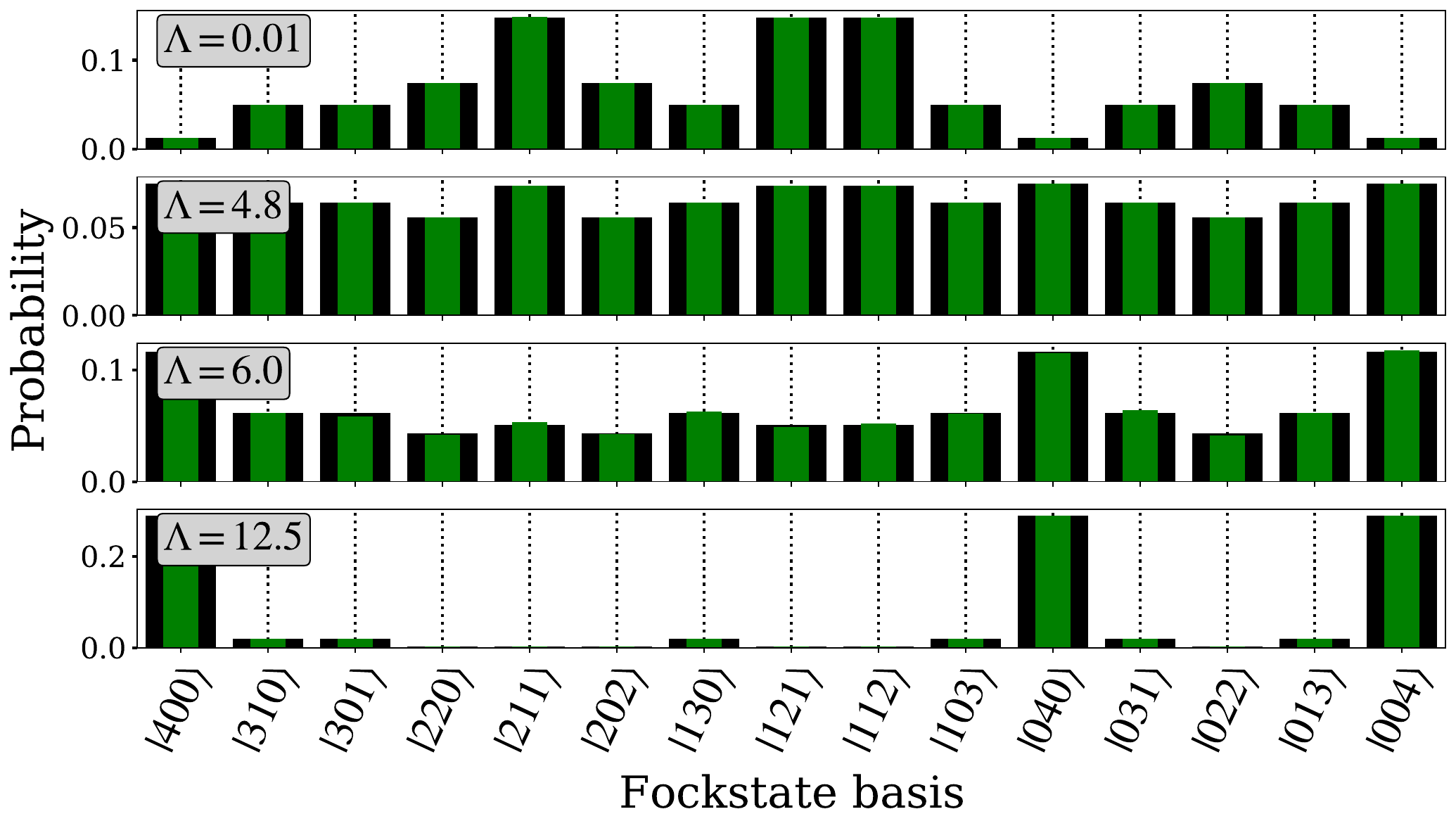}
    \caption{
    {
    \textbf{ Trial-state properties (3-site chain)}. Illustration of the properties of the quantum state obtained by minimization of the infidelity, using the BS-Kerr ansatz with $N_L=6$, $N_B=4$, and $\ket{\Psi_{ini}} = \ket{2,0,2}$. \textbf{Upper panel:} IPR and entanglement entropy $S$ of the resulting ansatz state (red dots) compared to the exact ground state (black curves) as a function of $\Lambda$. 
    \textbf{Lower panel: } Probability of occupation of the ansatz state in the Fock basis  (green bars) compared with exact diagonalization (black bars). The $\Lambda$ values chosen for the four configurations are marked in the upper panel with red dashed vertical lines (same as in Fig.~\ref{fig:3_4sites}).
    }
    }
    \label{fig:trimer_fidelity}
\end{figure}

\begin{figure}[t]
    \centering
    \includegraphics[width=\columnwidth]{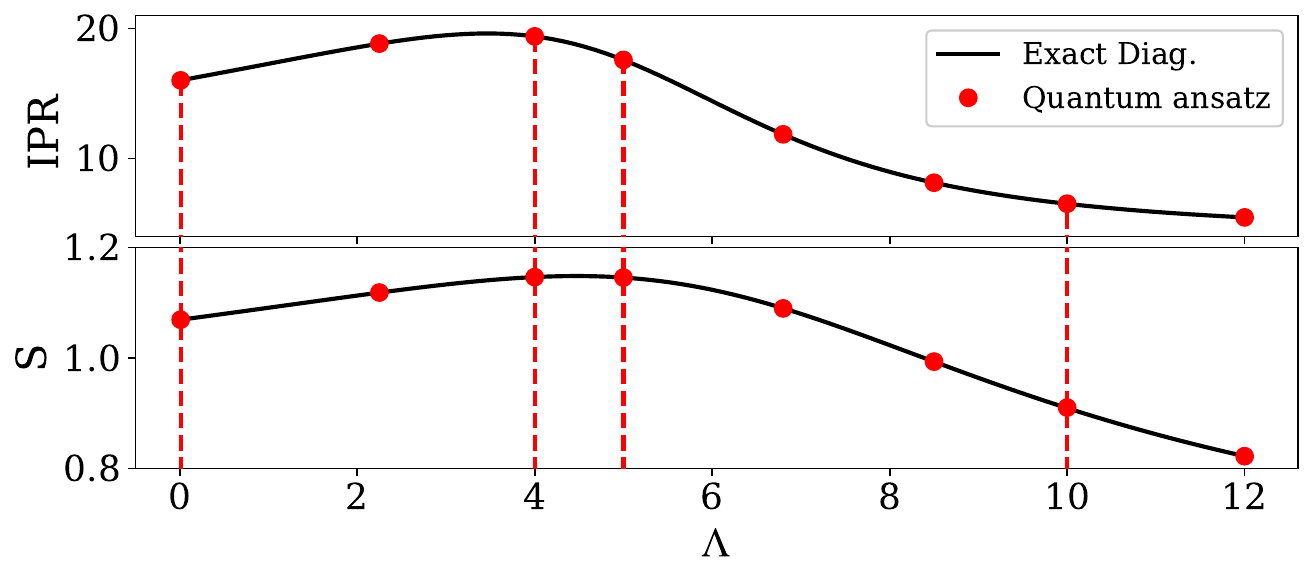}
    \includegraphics[width=\columnwidth]{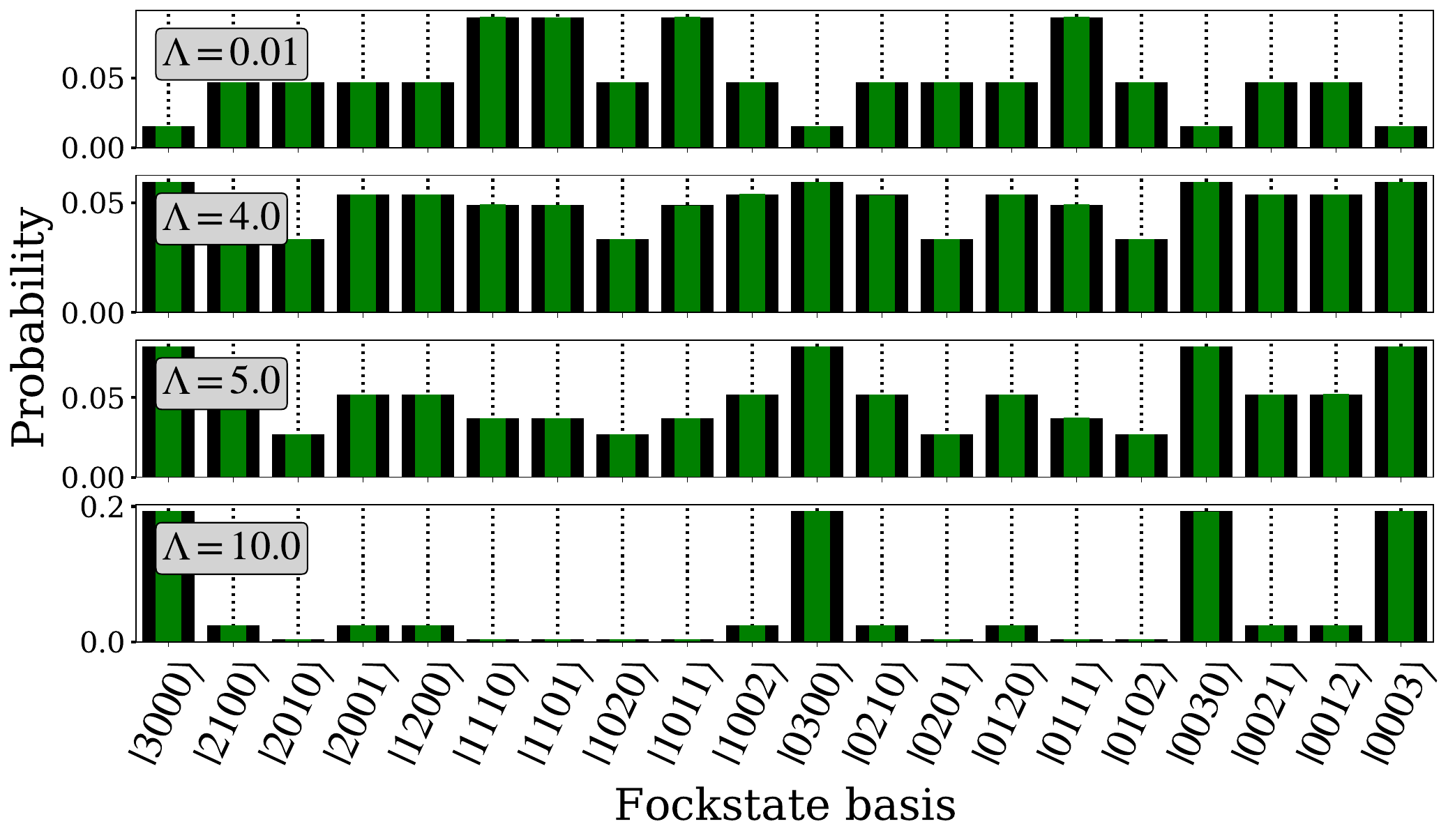}
    \caption{
    {
     \textbf{ Trial-state properties (4-site chain)}. Illustration of the properties of the quantum state obtained by minimization of the infidelity,
     using the BS-Kerr ansatz with $N_L=6$, $N_B=3$ and $\ket{\Psi_{ini}} = \ket{2,0,1,0}$. \textbf{Upper panel:} IPR and entanglement entropy $S$ of the resulting ansatz state (red dots) compared to the exact ground state (black curves) as a function of $\Lambda$. 
    \textbf{Lower panel: } Probability of occupation of the ansatz state in the Fock basis   (green bars) compared with exact diagonalization (black bars). The $\Lambda$ values chosen for the four configurations are marked in the upper panel with red dashed vertical lines (same as in Fig.~\ref{fig:3_4sites}).
    }
    }
    \label{fig:quadrimer_fidelity}
\end{figure}

The complexity grows even stronger for the four-site BH network.
We follow the exact same strategy than for the previous models by considering $\ket{\Psi_{ini}} = \ket{N_B,0,0,0}$ as the single-mode initial state and $\ket{\Psi_{ini}} =\ket{\frac{N_B}{2},0,\frac{N_B}{2},0}$ as the two-mode one (\textit{n.b.} for odd total number of bosons we took $\ket{\frac{N_B+1}{2},0,\frac{N_B+1}{2}-1,0}$ instead),
as well as $\Lambda$ values corresponding to different regimes of correlation marked by the vertical red-dashed lines in the right panel Fig.~\ref{fig:3_4sites}.
For this system, the dimension of the Fock space explodes very quickly
and fixing $N_B=5$ already leads to a number of bosonic configurations of $D_{\hat{\mathcal{H}}}=56$ that is larger than the size of the Fock space of the three-site model with $N_B=8$.
In our simulations, we chose to limit ourselves to a maximal number of bosons of $N_B=5$ due to the numerical complexity of the classical simulation of our VQA.
The performance of the BS-Kerr ansatz for $N_B=1$ to 5 is shown in Fig.~\ref{fig:Proof_4sites}.
In the full range of correlation regimes, the photonic circuit is again able to encode the ground state with high fidelity for a circuit depth of $N_L \approx 2 N_B$. {Here again we see that the single-mode initial state seems to perform better compared to the two-mode case. The associated maximum number of layers being $N_L=8$ for the latter (with 32 Kerr gates) compared to $N_L=10$ for a two-mode initial state (with 40 Kerr gates). We also retrieve on Fig.~\ref{fig:Proof_4sites} the case of an exact encoding with one layer for the initial single-mode state for weakly correlated regime $\Lambda = 0.01$.} 

\begin{figure*}
    \centering
    \includegraphics[width=\textwidth]{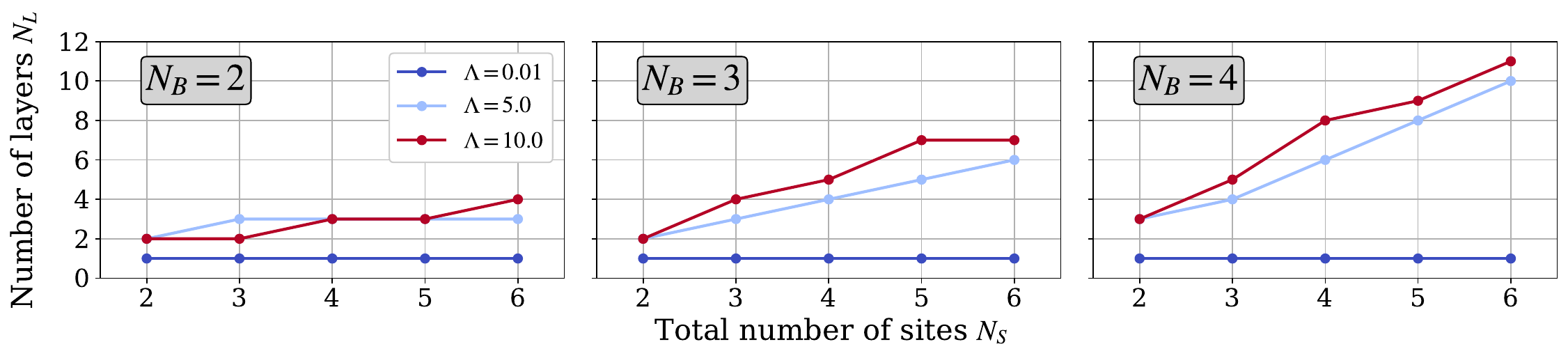}
    \caption{\textbf{Scaling of the ansatz encoding capacity with the system's size $\boldsymbol{N_S}$.} Minimum depth required (in terms of number of layers $N_L$) to generate a fidelity $\mathcal{F} \geq 99\%$ [see Eq.~(\ref{eq:fidelity})] with the minimal BS-Kerr ansatz as a function of the number of sites (ranging from $N_S=2$ to 6). Here, three different values of many-body interaction ($\Lambda=0.01$, 5 and 10) are considered to span the different regimes of correlation. The three panels, from left to right, show results obtained with a maximum number of $N_B=2$, 3 and 4 bosons, respectively. A single-mode initial state is employed in all simulations.}
    \label{fig:N_S_dependency}
\end{figure*}

{
As an addition, Figs.~\ref{fig:trimer_fidelity} and \ref{fig:quadrimer_fidelity} illustrate the structural transition of the BH ground state for the three and four-site BH networks, respectively. Similarly to Fig.~\ref{fig:dimer_fidelity}, the upper part of the plots show the evolution of the IPR and the entanglement entropy of trial states (red dots) compared to exact ground state (black curves) as a function of $\Lambda$. Note that, all the trial states generated here present a fidelity of $\mathcal{F} \geq 99\%$. The lower panels show the density of probability of the trial states (green bar) compared to the exact ground state (black bars).
Just as in the BH dimer, the IPR, the entanglement entropy $S$ and the density of probability
of the resulting trial states are in very good agreement with the exact ground state.
}

{
As a summary of the results presented so far, a remarkable property is that very high fidelity encoding can be performed with the minimal ansatz for the tree system, whatever the regimes of many-body interactions. Globally, our results suggest that, for a fixed system size, the number of layers required to ensure an encoding of $\mathcal{F} \geq 99 \%$ scales linearly with the number of bosons. Furthermore, our simulations also demonstrate that high fidelity may be obtained even if we consider different initial states $\ket{\Psi_{ini}}$ that are simple Fock states (and not complex quantum superpositions of many Fock states). 
Hence, the minimal ansatz is expressive enough to accurately transform such single Fock states into complex linear combinations of $D_{\hat{\mathcal{H}}}$ configurations encoding strong quantum correlations.
}


\subsubsection{ Effects of the system size $N_S$ }\label{sec:scaling_NS}

{
Let us now investigate how the increase in the system size can affect the expressibility and the resource efficiency of the minimal BS-Kerr ansatz.
In analogy to the previous study, we evaluate the number of layers $N_L$ required to satisfy $\mathcal{F} = 99\%$, but as a function of the total number of sites $N_S$ rather than the number of bosons $N_B$. 
The results are presented in Fig.~\ref{fig:N_S_dependency}
for three values of interaction parameter ($\Lambda=0.01$, 5 and 10)
and a number of bosons $N_B=2$, 3 and 4.
As readily seen in Fig.~\ref{fig:N_S_dependency}, the number of layers $N_L$ required to ensure a high fidelity scales linearly with the number of sites of the BH system (whatever the number of bosons $N_B$ is). 
Focusing on the strong many-body interaction regime (i.e. $\Lambda=10$),
a scaling of $N_L \sim 0.5 N_S $ is obtained for $N_B=2$ bosons, $N_L \sim 1.5 N_S $ for $N_B=3$ bosons, and $N_L \sim 2 N_S $ for $N_B=4$ bosons. 
The  prefactor in the linear scaling seems to slightly increase with the total number of bosons $N_B$ (a trend already observed Figs.~\ref{fig:dimer_fidelity}, \ref{fig:trimer_fidelity} and \ref{fig:quadrimer_fidelity}). 
Note that such linear behaviors strongly contrast with the non-linear increase of the number of configurations (i.e.~the size $D_{\hat{\mathcal{H}}}$ of the Fock space) necessary to represent the exact ground-state of the BH model.
Indeed, as an illustration, note that in the case of $N_B=4$ bosons the number of configurations involved starts from $D_{\hat{\mathcal{H}}}=5$ for the dimer ($N_S=2$) to reach $D_{\hat{\mathcal{H}}}=126$ for the BH chain with $N_S=6$ sites (which represents the maximum number of Fock states generated in our study).  
On another note  about the weakly correlated regime (dark blue curves in Fig.~\ref{fig:N_S_dependency}), we see here again that a single layer is sufficient to ensure a high fidelity encoding of the BH ground state, for all the number of sites $N_S$ and number of bosons $N_B$ considered (see Appendix~\ref{app:demo} for a mathematical proof).
Finally, to summarize our observations realized so far, our results from the previous sections combined with the ones presented here reveal that the number of layer $N_L$ required to accurately encode the ground state of the BH model presents a linear trend with respect to both $N_B$ and $N_S$. This being verified at least for the small to medium-sized BH systems considered herein. }


 \subsection{Capacity of the interferometer-Kerr ansatz}\label{subsec:interferometer_res}

\begin{figure}[t]
    \centering 
    \includegraphics[width=\columnwidth]{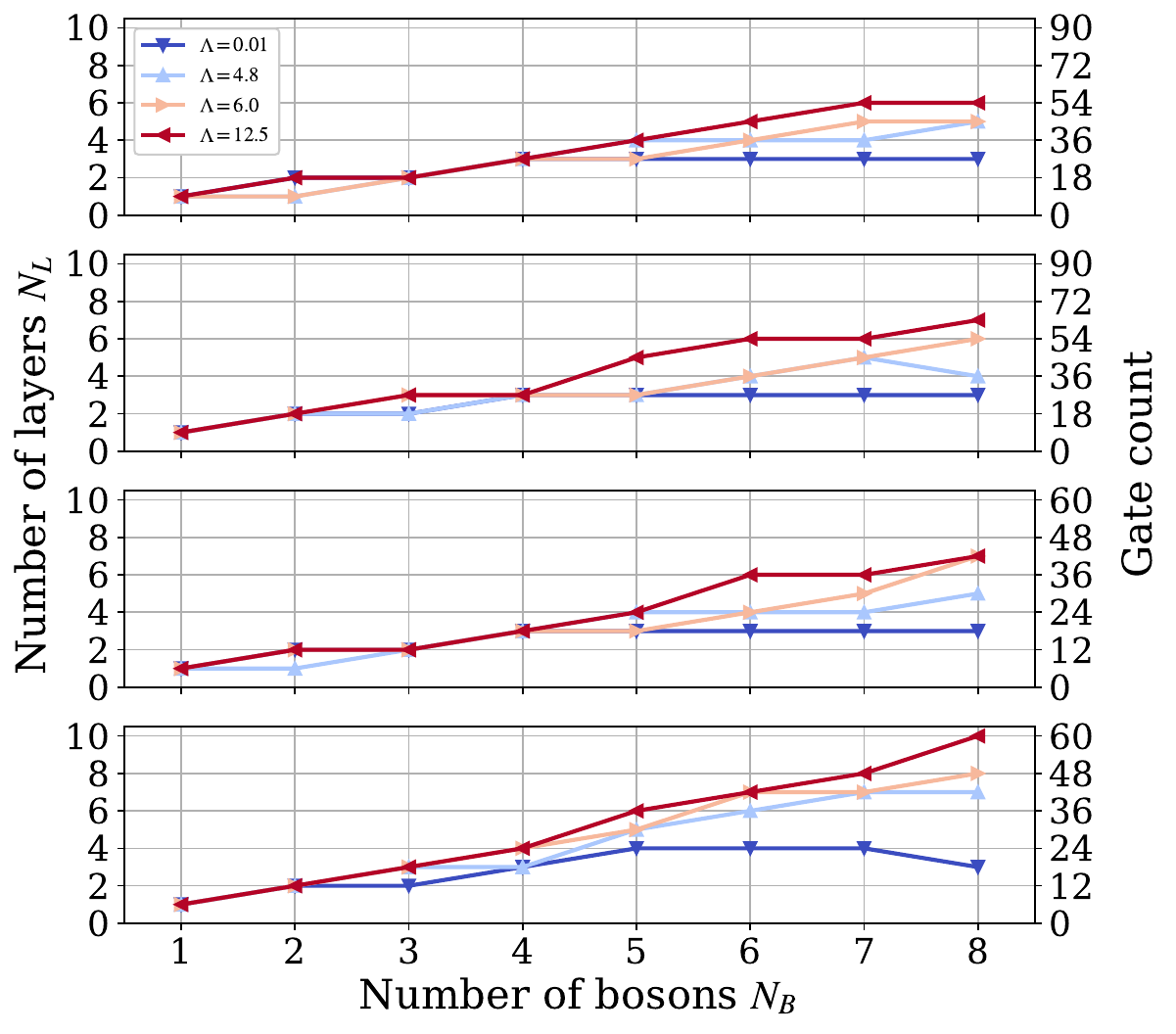}
    \caption{\textbf{Ansatz encoding capacity (3-site chain).} Minimum depth required (in terms of number of layers $N_L$ and gate count) to generate a fidelity $\mathcal{F} \geq 99\%$ [see Eq.~(\ref{eq:fidelity})] for different values of $\Lambda$ and $N_B$, using the two-mode initial state. \textbf{First panel:} Full ansatz, \textbf{Second panel:} Phase angles of the beam-splitters are set to 0, \textbf{Third panel:} Phase angles of the beam-splitter are reintroduced but rotation gates are removed, \textbf{Fourth panel:} Phase angles of the beam-splitters are set to 0 and the rotation gates are removed.}
    \label{fig:Proof_3sites_interferometerKerr}
\end{figure}

Let us now turn to the interferometer-Kerr ansatz described in Sec.~\ref{sec:intK_ansatz}.
By using an interferometer composed of more beam-splitter and rotation gates,
the interferometer-Kerr ansatz is expected to perform as efficiently as the BS-Kerr ansatz but with a reduced number of layers, hence a reduced number of Kerr gates.
As readily seen in the top panel of Fig.~\ref{fig:Proof_3sites_interferometerKerr}, only $N_L = 6$ layers
are necessary to encode more than 99$\%$ of the ground state of the three-site model for $N_B=8$, compared to the 10 layers required by the BS-Kerr ansatz (see Fig.~\ref{fig:Proof_3sites}).
Note that reducing the number of layers in the interferometer-Kerr ansatz does not necessarily reduce the total number of gates compared to the BS-Kerr ansatz.
Indeed, $N_L = 6$ layers in the interferometer-Kerr ansatz is equivalent to 54 gates, while $N_L = 10$ for the BS-Kerr ansatz gives 50 gates.
However, the number of Kerr gates is always directly proportional to the number of layers and the number of modes as $N_{\rm Kerr} = N_L\times N_S$ such that the number of Kerr gates is almost reduced by half when using the interferometer-Kerr ansatz (designed for this exact purpose).
Note that the phase angles of the beam-splitter gates were not set to 0 compared to the BS-Kerr ansatz.
For a more fair comparison, we set them to 0 in the second panel of Fig.~\ref{fig:Proof_3sites_interferometerKerr}.
One can notice a slight increase in the number of layers, especially when the correlation strength $\Lambda$ increases, but only up to $N_L = 7$.
Hence, for the three-site model, the interferometer-Kerr ansatz guaranties an efficient encoding of the ground state
with $N_L \sim 0.8 N_B$ compared to $N_L \sim 1.25 N_B$ for the BS-Kerr ansatz, for all values of $\Lambda$.

\begin{figure}[t]
    \centering 
    \includegraphics[width=\columnwidth]{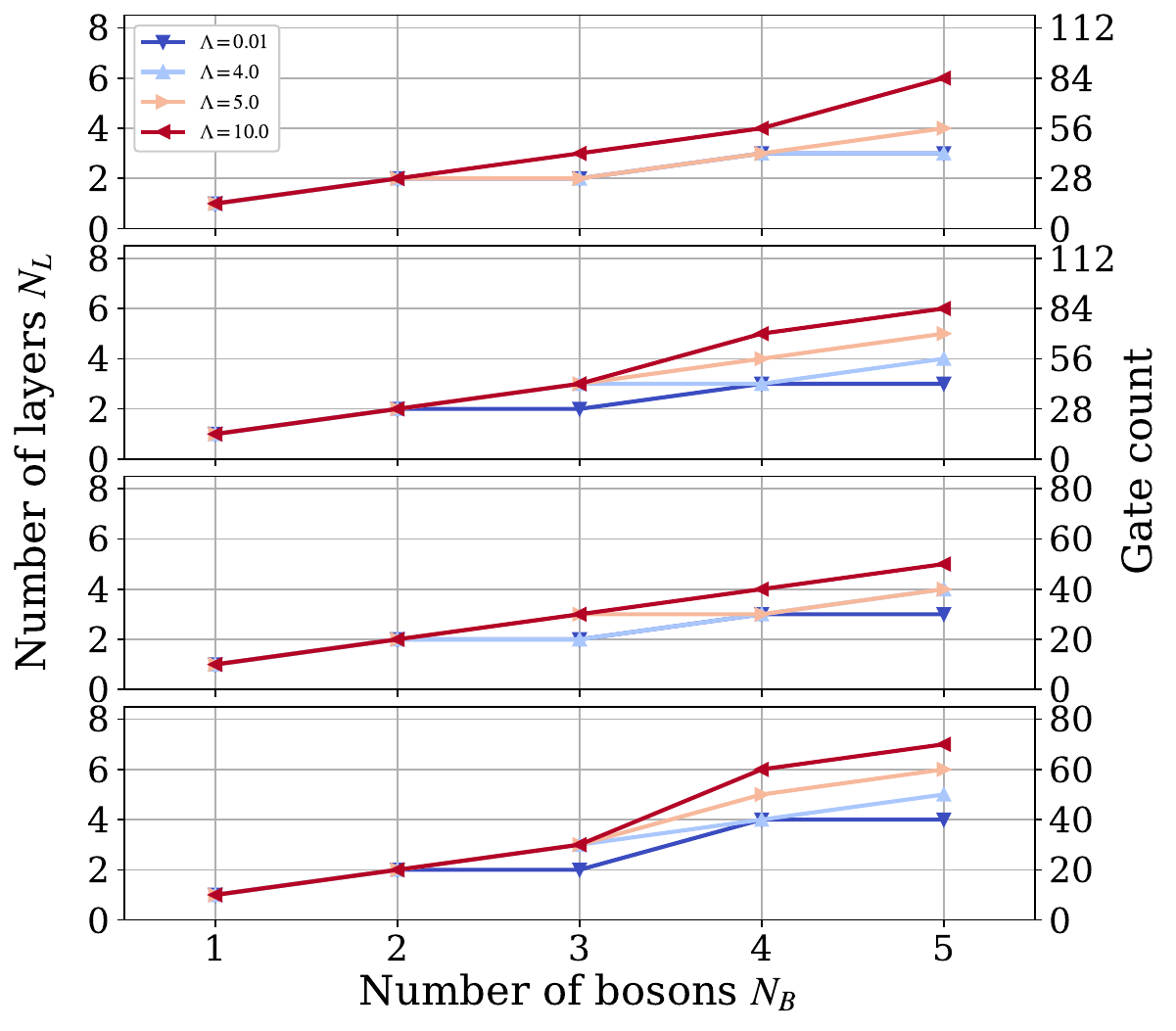}
    \caption{\textbf{Ansatz encoding capacity (4-site chain).} Minimum depth required (in terms of number of layers $N_L$ and gate count) to generate a fidelity $\mathcal{F} \geq 99\%$ [see Eq.~(\ref{eq:fidelity})] for different values of $\Lambda$ and $N_B$, using the two-mode initial state. \textbf{First panel:} Full ansatz, \textbf{Second panel:} Phase angles of the beam-splitters are set to 0, \textbf{Third panel:} Phase angles of the beam-splitter are reintroduced but rotation gates are removed, \textbf{Fourth panel:} Phase angles of the beam-splitters are set to 0 and the rotation gates are removed.}
    \label{fig:Proof_4sites_interferometerKerr}
\end{figure}

We just saw that the total number of gates of the interferometer-Kerr ansatz can still be higher than of the BS-Kerr ansatz, despite a significant reduction in the number of Kerr gates.
As mentioned in Ref.~\cite{clements2016optimal}, the rotation gates of the interferometer are physically irrelevant for most applications.
These gates are removed from the ansatz in the third and fourth panels of Fig.~\ref{fig:Proof_3sites_interferometerKerr}, where the phase angles of the beam-splitter are either free or set to 0, respectively.
Looking at the fourth panel, it is clear that simultaneously removing the phase angles of the beam-splitter gates and the rotation gates deteriorates the performance of the ansatz, and does not lead to any improvement over the BS-Kerr ansatz.
Even for $\Lambda = 0.01$, this is the only case where more than $N_L = 3$ layers are required to accurately encode the ground state.
However, removing the rotation gates while keeping the phase angle of the beam-splitter gates (third panel) {gives similar results than} doing the exact opposite, i.e. keeping the rotation gates and removing the phase angle of the beam-splitter gates (second panel).
By doing so, the number of parameters remains the same, but the total number of gates can be reduced, for instance from 63 to 42 for $N_B = 8$.
{However, for larger systems, there are more additional phase angles than rotation gates as the interferometer is composed of $N_S(N_S - 1)/2$ beam-splitters but only $N_S$ rotation gates, and getting rid of one or the other might lead to different results.
As there are more beam-splitters than rotation gates, we expect that getting rid of the rotation gates is safe, especially as they are supposed to be physically irrelevant for most applications~\cite{clements2016optimal}.}


Turning to the four-site model in Fig.~\ref{fig:Proof_4sites_interferometerKerr}, we reach the exact same general conclusions as for the three-site model.
It seems that the circuit depth required to encode the ground state of this model tends to $N_L \sim 1.2 N_B$, which again improves over $N_L \sim 2 N_B$ for the BS-Kerr ansatz.
Without the rotations and the phase angles (fourth panel), the number of layers tends to increase in all regimes of correlation, while keeping the phase angles and removing the rotation gates seems to provide the best trade-off between the total number of gates, the number of Kerr gates (proportional to the number of layers), and the number of parameters.

\section{VQE simulations}\label{sec:VQE}


In the previous section, we gave numerical evidence on the capacity of our proposed ansatze to encode the ground state of the attractive BH model.
In this section, we switch to more practical calculations by simulating VQE protocols where the cost function to minimize is the energy of the system,
\begin{equation}
    E(\vec{\theta}) = \bra{\Psi_{ini}} \hat{U}(\vec{\theta}) ^\dagger \hat{\mathcal{H}} \hat{U}(\vec{\theta}) \ket{\Psi_{ini}} \geq E_0. \label{eq:energy_estimation}
\end{equation}
For the sake of conciseness, we focus on the BS-Kerr ansatz only.
First, we briefly demonstrate the accuracy of the results obtained on the three and four-site models. 
Second, we use the BH dimer as a test-bed to tackle the practical simulation of a VQE experiment on a CV device.
This second part will be the occasion to compare the results obtained with an ideal realization of a VQE to a more realistic one, where the estimation of the energy $E(\vec{\theta})$ is based on a photon-counting protocol.

\subsection{Ideal VQE simulations for the three- and four-site BH model}\label{subsec:VQE_34}

\begin{figure}[t]
    \centering 
    \includegraphics[width=\columnwidth]{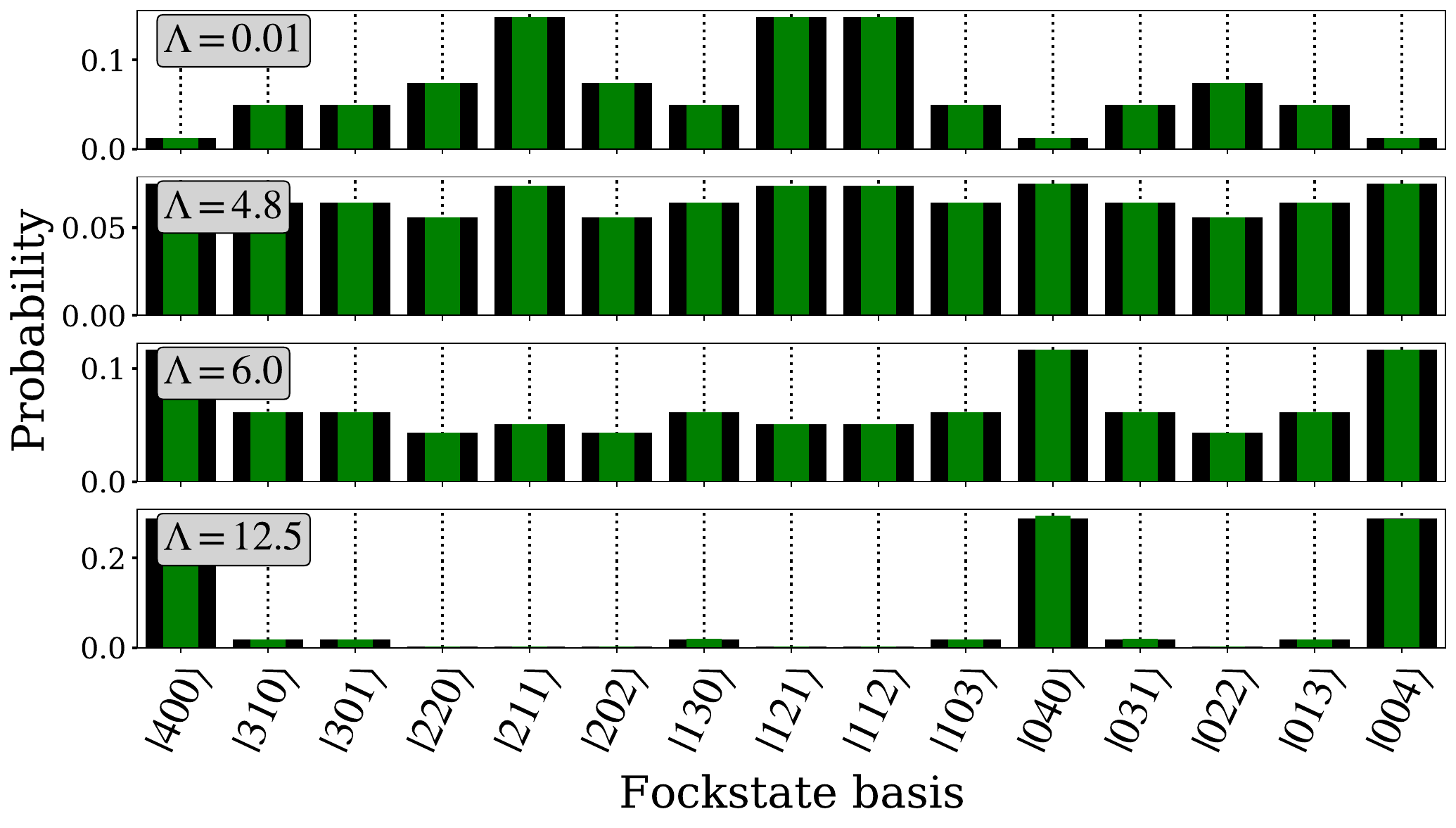}
    \includegraphics[width=\columnwidth]{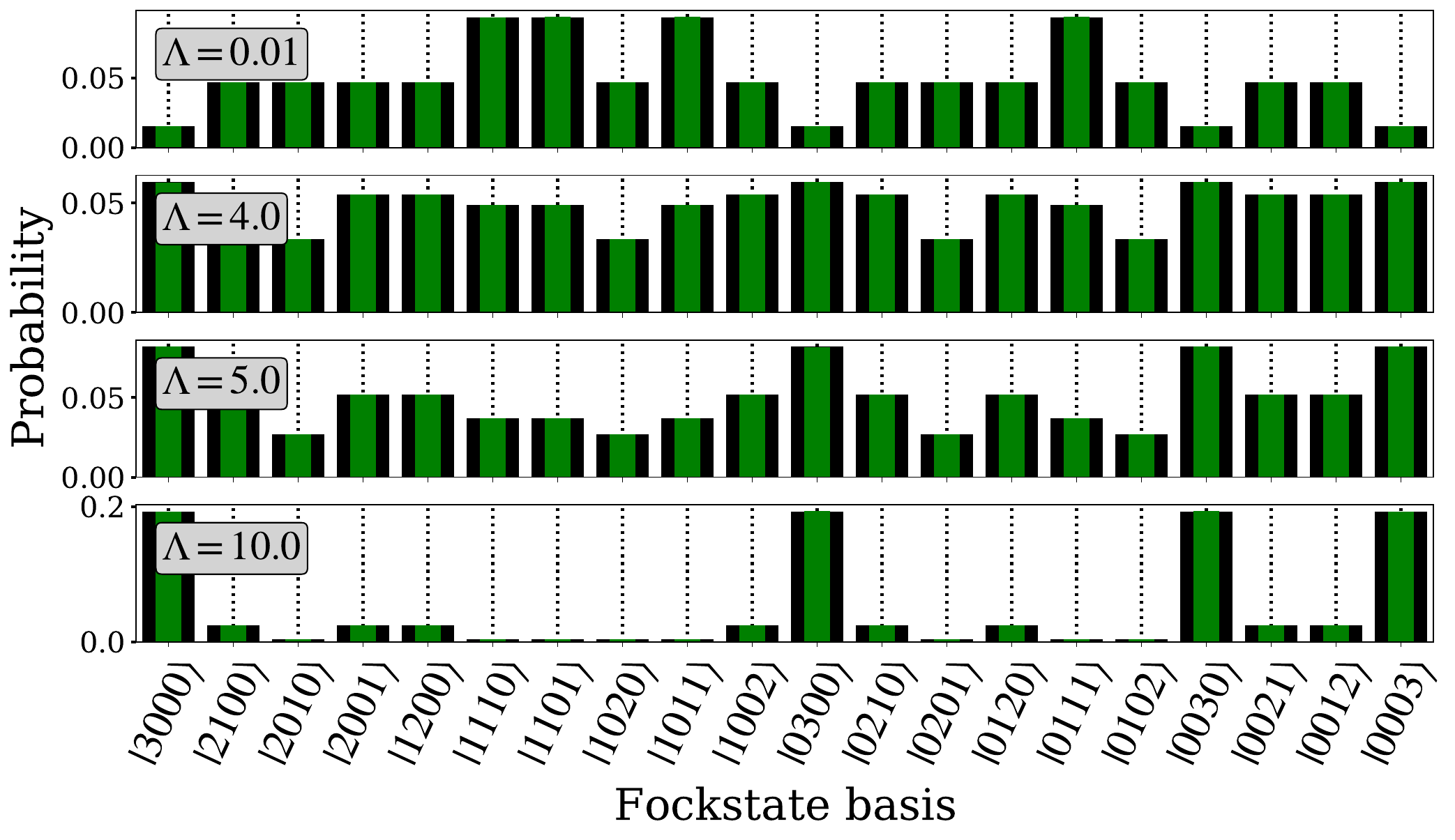}
    \caption{ \textbf{ Trial-state properties (3-/4-site chains)}.
    Probability of occupation in the Fock basis obtained by ED (black bars) and VQE simulations,
    using the BS-Kerr ansatz (green bars).
    All parameters were initialized with random amplitudes drawn in $[-0.05,+0.05]$, and the BFGS optimizer with a maximum number of 2000 iterations was used.
    \textbf{Upper panel:} Ground state of the three-site model with $N_B=4$, $N_L=6$ and $\ket{\Psi_{ini}} = \ket{2,0,2}$
    \textbf{Lower panel:} Ground state of the four-site model with $N_B=3$, $N_L=6$ and $\ket{\Psi_{ini}} = \ket{2,0,1,0}$.
   }
    \label{fig:VQE_34}
\end{figure}

In this section, ideal simulations of the VQE algorithm are realized (i.e.~infinite sampling, no photon loss, no error in the circuit\ldots). 
This means that we assume that we can extract the trial state $\ket{\Psi(\vec{\theta}} = \hat{U}(\vec{\theta}) \ket{\Psi_{ini}} $ at the end of the circuit, which makes it possible to exactly estimate the energy $E(\vec{\theta})$ during the optimization process.

Figure~\ref{fig:VQE_34} shows 
the probability of occupation in the Fock basis of the ground states
obtained at the end of the VQE algorithm for different $\Lambda$ values (consistent with the simulations realized in the previous section).
Note that $N_L=6$ layers and initial two-mode states ($\ket{\Psi_{ini}} = \ket{2,0,2}$ and $\ket{\Psi_{ini}} = \ket{2,0,1,0}$) are considered. 
The depth of the circuits imposes to optimize $30$ (respectively $42$) parameters for the three-site (four-site) network.

As shown in Fig.~\ref{fig:VQE_34}, the probability of occupation follows the same trend as the previous ones in Figs.~\ref{fig:trimer_fidelity} and \ref{fig:quadrimer_fidelity}, where the infidelity $\mathcal{I}$ was minimized instead of the ground-state energy. 
In all cases of Fig.~\ref{fig:VQE_34}, the resulting states have fidelity $\mathcal{F}\geq 99\%$ and an energy error $\Delta E  = E(\vec{\theta}^*) - E_0 \leq 10^{-5}J$ (where $\vec{\theta}^*$ denotes the parameters that minimize the cost function), thus showing an excellent convergence of the VQE algorithm. 
Note that similar ideal VQE simulations have been also performed
using single-mode initial states (i.e.~$\ket{\Psi_{ini}} = \ket{4,0,0}$ and $\ket{\Psi_{ini}} = \ket{3,0,0,0}$, not shown),
and led to similar accuracy. These results give a good
demonstration of the ability of the BS-Kerr ansatz to map different initial states to the exact ground state of the attractive BH model (in all correlation regimes considered herein) in a more practical VQE context.

\subsection{Ideal VQE v.s. realistic VQE with sampling noise: application to the BH dimer.}\label{subsec:VQE_noise}

Let us now turn to the implementation of a more realistic execution of the VQE algorithm, as it could be done in a practical experiment.

\subsubsection{Measure of the expectation value of the Hamiltonian by a photon counting approach }\label{sec:measurements}

In practice, a realistic simulation of a VQE algorithm requires to measure the energy of a given trial state at the end of the quantum circuit.
{ For this purpose, we propose a measurement protocol based on photon counting method which can be realized with photon number resolving (PNR) detectors~\cite{jonsson2019evaluating,provaznik2020benchmarking}. Note that the development of PNR (and other similar techniques) is critical for the implementation of scalable quantum information processing (see the review~\cite{slussarenko2019photonic} and references therein about photon detection methods for photonic quantum computing).  These methods are still actively investigated to push the limit of maximum number of detectable photons, with a current maximum of around several tenths of photons that are resolved simultaneously.}

In the following, the notation $\langle \hat{O} \rangle_\Psi \equiv \bra{\Psi } \hat{O} \ket{\Psi }$ refers to the quantum average of an operator $\hat{O}$ realized at the end of the circuit when the photons are in the final state $\ket{\Psi}$. Starting from the BH Hamiltonian $\hat{\mathcal{H}}$ defined in Eq.~(\ref{eq:HBH}),
measuring the energy of the final photonic state $\ket{\Psi}$ of the circuit requires measuring two different contributions,
\begin{equation}\label{eq:onebody_term}
    \langle \hat{\mathcal{H}}_\text{Hoppings} \rangle  =  -J \sum_{\langle p,q\rangle}^{N_S} \langle  (a_p^\dagger a_q + a_q^\dagger a_p ) \rangle_\Psi,
\end{equation}
and
\begin{equation}\label{eq:twobody_term}
   \langle  \hat{\mathcal{H}}_\text{Loc-Int}  \rangle = -\frac{U}{2} \sum_p^{N_S}  \langle n_p(n_p-1) \rangle_\Psi.
\end{equation}
The one-body (hopping) term in Eq.~(\ref{eq:onebody_term}) can be measured
by introducing additional 50/50 beam-splitter gates (i.e.~with $\phi=0$ and $\theta=\pi/4$) between each pair of modes before measuring, thus leading to
\begin{equation}
B_{p,q}(\pi/4,0)\left[a_p^\dagger a_q + a_q^\dagger a_p \right]B_{p,q}(\pi/4,0)^\dagger = n_q-n_p
\end{equation}
such that
\begin{equation}
    \langle  a_p^\dagger a_q + a_q^\dagger a_p  \rangle_\Psi = \langle n_q-n_p \rangle_{\tilde{\Psi}} 
\end{equation}
where $\ket{\tilde{\Psi}} = B_{p,q}(\pi/4,0)\ket{\Psi}$.
Thus, evaluating the hopping term between two modes with respect to the original trial state $\ket{\Psi}$ is equivalent to measuring the difference of average photon numbers in the two same modes after applying 50/50 beam-splitter.

Turning to the many-body interaction term in Eq.~(\ref{eq:twobody_term}), 
note that it is directly related to the photon statistics of each individual mode at the end of the circuit:
\begin{equation}
\label{eq:locint}
    \langle n_p(n_p-1) \rangle_\Psi = \text{Var}(n_p)_\Psi +  \langle n_p \rangle_\Psi^2 - \langle n_p \rangle_\Psi,
\end{equation}
where $\text{Var}(n_p) = \langle n_p^2 \rangle_\Psi - \langle n_p \rangle^2_\Psi$ is the variance of the photon number in mode $p$ at the end of the circuit.
In other words, to estimate the contribution of the many-body interaction term to the total energy, we only need to evaluate (over many samples) the first two moments of the number operator for each mode $p$.

Although not studied in this work, an extended version of the BH model -- originally designed to study the supersolid phase of helium~\cite{matsuda1970off,liu1973quantum} -- is sometimes considered (see for example Refs.~\cite{chen2018quantum,baier2016extended,rossini2012phase,ohgoe2012ground}) and reads
\begin{equation}\label{eq:extendedBH}
    \hat{\mathcal{H}}'  =  \hat{\mathcal{H}} + \sum_p \mu_p n_p + \sum_{p,q} V_{pq} n_p n_q,
\end{equation}
where $\mu_p$ is usually referred to as
the ``chemical potential'', while the last term encodes the dipole-dipole interaction of two packets of bosons localized on two different sites $p$ and $q$.
The expectation value of these two additional terms can also be estimated via photon counting. 
First, the expectation value of the chemical potential
\begin{equation}
    \langle \hat{\mathcal{H}}_\text{Chem-Pot} \rangle_\Psi = \sum_p \mu_p \langle n_p \rangle_\Psi
\end{equation}
is relatively simple to estimate
as it only requires to do local photon-counting on each mode. 
The energy contribution of the dipole-dipole term
 \begin{equation}
   \langle \hat{\mathcal{H}}_\text{dip-dip} \rangle_\Psi=   \sum_{p,q} V_{pq} \langle n_p n_q \rangle
\end{equation}
can be found by evaluating the covariance matrix of the photon number distribution in all the modes:
\begin{align}\label{eq:cov}
    \langle n_p n_q\rangle = \mathrm{Cov}(n_p,n_q) + \langle n_p\rangle\langle n_q\rangle.
\end{align}
Note that setting $p=q$ in Eq.~(\ref{eq:cov})
gives back Eq.~\eqref{eq:locint}.

Note that the extended Bose--Hubbard model has been shown to capture essential properties of many systems~\cite{lin2017disordered}, including ultracold atoms and molecules in optical lattices~\cite{goral2002quantum,buchler2003supersolid,scarola2005quantum,scarola2006searching,bloch2008many,baier2016extended}, Josephson junction arrays~\cite{van1995quantum,roddick1995supersolid}, and narrow-band superconductors~\cite{micnas1990superconductivity}.
Testing our proposed ansatze on this model is a natural follow-up of this study and is left for future work.

\subsubsection{VQE with and without sampling noise}

\begin{figure*}
    \centering
    \includegraphics[width=0.49\textwidth]{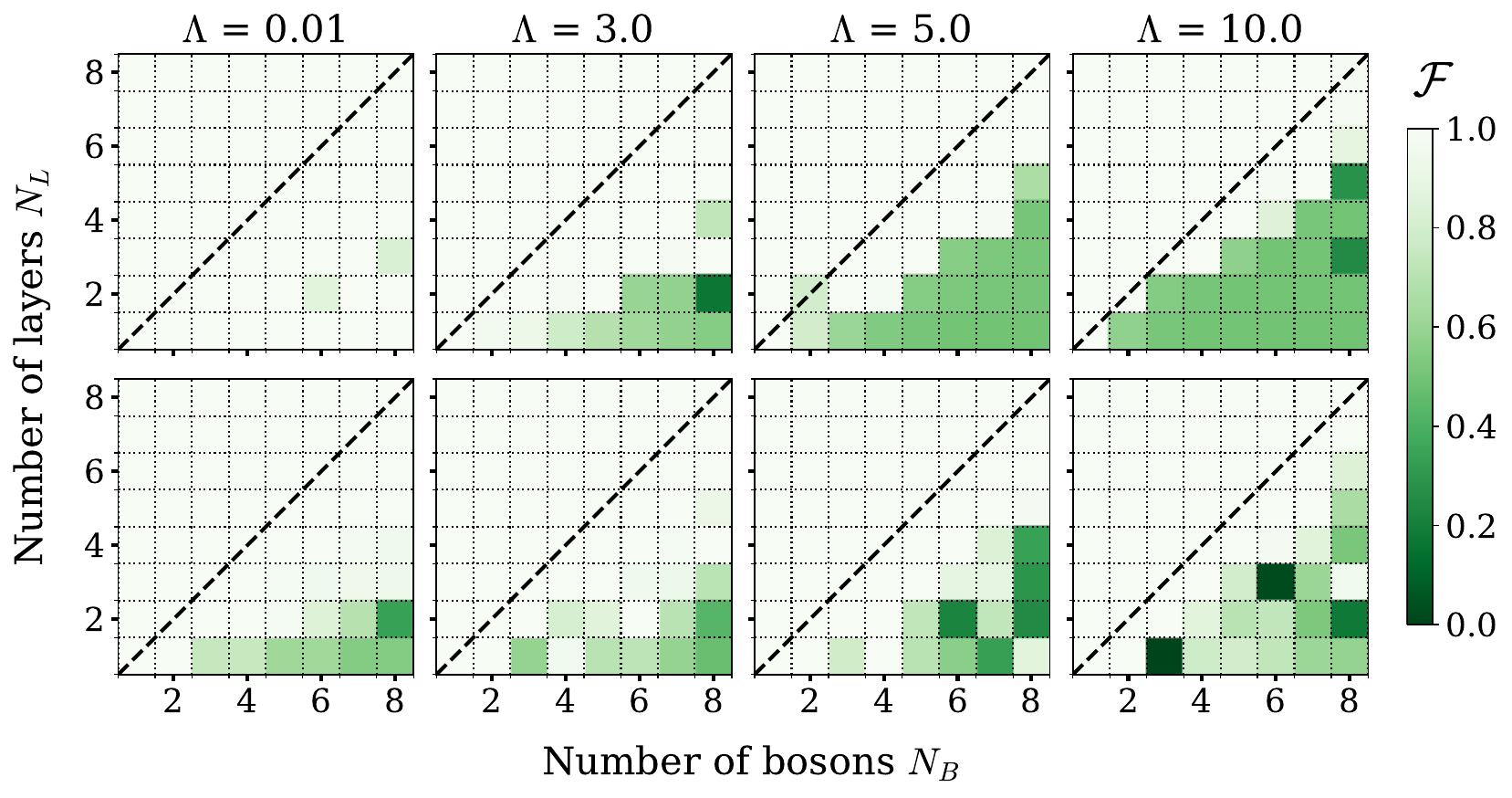}
    \includegraphics[width=0.49\textwidth]{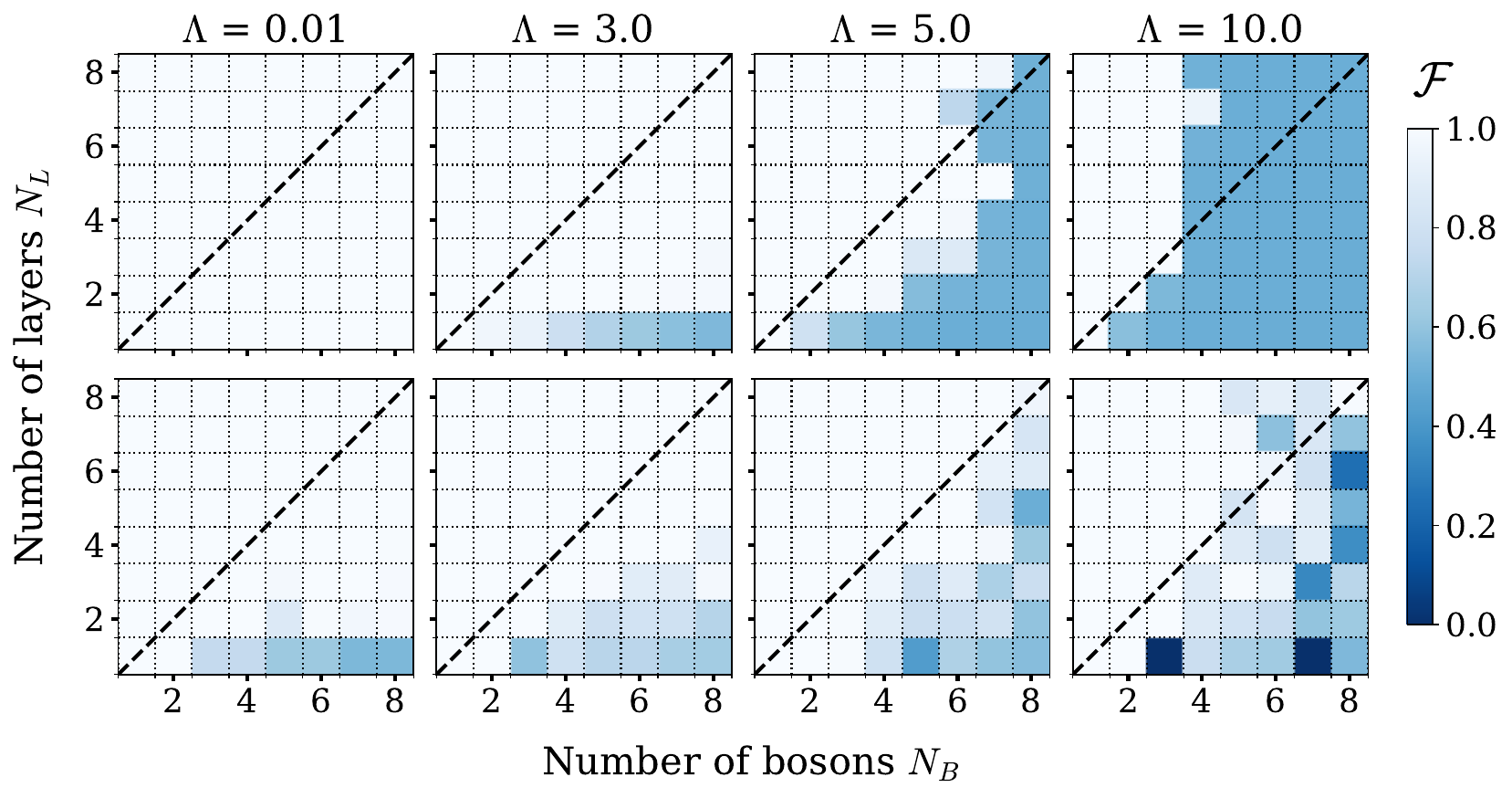}
    \caption{   
    \textbf{Ground-state fidelity $\mathcal{F}$ of trial states obtained with VQE simulations for the BH dimer (using the BS-Kerr ansatz).} The final fidelity is shown for four different values of $\Lambda$ and as a function of the number of bosons $N_B$ and the number of layers $N_L$ in the circuit. 
    \textbf{Left panel:} noiseless VQE simulation using the BFGS optimizer with a maximum number of 2500 iterations. All parameters are initialized with a random amplitude in $[-1,+1]$. 
    \textbf{Right panel:} VQE simulation with sampling noise using the noise-robust optimizer CMA-ES~\cite{hansen2016cma} with an initial step-size parameter of $\sigma_0=0.05$ and a maximum number of 20000 functions evaluations (each function evaluation stems from sampling of $10^8$ samples of the quantum circuit). All parameters are initialized with a random amplitude in $[-0.1,+0.1]$.
    On each panel, upper and lower rows show results obtained by starting from a single-mode and a two-mode initial state, respectively.
    }
    \label{fig:fidelity_map}
\end{figure*}

Focusing on the BH dimer, we now investigate how the sampling of the energy $E(\vec{\theta})$ can affect the quality of the VQE results. 
Compared to the ideal VQE simulations realized above, considering sampling noise leads to fluctuations in the energy $E(\vec{\theta})$ that makes the optimization harder in practice.

Fig.~\ref{fig:fidelity_map}
shows the ground-state fidelity $\mathcal{F}$ obtained from VQE simulations with (right panel in blue) and without (left panel in green) sampling noise.
In the latter case (with sampling noise),
the energy is calculated using our photon-counting protocol described in Sec.~\ref{sec:measurements}.
To mitigate the impact of noise in the classical optimization, we employed a ``noise-proof'' optimizer, namely CMA-ES~\cite{hansen2016cma}.
The number of samples is fixed to $10^8$
and photon-counting is only realized on the first mode of the circuit, as the conservation of the total number of bosons imposes $N_B = n_1 + n_2$ (i.e.~measuring the number of photons $n_1$ in the first mode gives directly the number of photons $n_2$ in the second mode, assuming no photon loss).

Let us first focus on the ideal VQE simulation (left panel of Fig.~\ref{fig:fidelity_map}). In analogy with the previous study in Sec.~\ref{sec:num_results_expressibility},
the ground state is encoded with high fidelity
whatever the initial state and correlation regime considered.
In all of these cases, a sub-linear limit marks the separation between the regions of high and low fidelity-encoding (in average $N_L \approx 0.6 N_B $).
The slope of this limit actually grows with the strength of $\Lambda$, thus indicating that ``cat-like'' states that arise for strong many-body interactions can be the more complex case to encode with a VQE using the BS-Kerr ansatz.
 
Turning to the noisy VQE simulation (right panel of Fig.~\ref{fig:fidelity_map}) based on our photon-counting protocol,
one can see that the resulting ground-state fidelity
strongly depends on the nature of the initial state. 
Indeed, for $\Lambda \leq 5$, the results obtained with sampling noise are globally comparable to the ones from an ideal VQE.\linebreak
Interestingly, when the single-mode initial state is used, an even better fidelity is reached in this regime of weak correlation.
However, for $\Lambda=5$, the resulting fidelities deteriorate for both initial states, especially when the number of bosons increases. The deterioration is even more important when considering the strong interaction regime $\Lambda = 10 $.
In this case, a vertical limit separates the high- and low-fidelity regions at $N_B = 4$ when the single-mode initial state is used. 
{Over this limit, the fidelity of the VQE trial state gets stuck at $\mathcal{F}\sim 50\%$. This occurs when the many-body interaction strongly intensifies, which leads to the ground state and first excited state to be quasi degenerate. This phenomenon arises as the many-body attraction essentially favors configurations that gather as many bosons as possible on a same site (as illustrated in Fig.~9, the most favorable configurations are $\ket{N_B,0}$ and $\ket{0,N_B}$, followed by $\ket{N_B-1,1}$ and $\ket{1,N_B-1}$, and then $\ket{N_B-2,2}$ and $\ket{2,N_B-2}$, etc.).
If we simplify by considering the most favorable configurations $\ket{N_B,0}$ and $\ket{0,N_B}$, the resulting quasi-degenerate ground and first excited state can roughly be seen as a ``plus version'' of cat-state $\ket{\Psi_0} \propto  \ket{N_B,0} + \ket{0,N_B}$ and a ``minus version'' $\ket{\Psi_1} \propto \ket{N_B,0} - \ket{0,N_B}$, respectively. In this case, running a noisy VQE simulation with an initial single-mode state $\ket{\Psi_{ini}} = \ket{N_B,0}$ produces a final energy that stands in between both quasi degenerate states' energies (not shown here).  As the sampling noise typically equates the amplitude of the spectral gap between both states, the minimization gets stuck here. The final fidelity of the trial states is then of $50\%$ with both quasi degenerate states, indicating then that $\ket{\Psi(\theta)}$ turned into a linear combination of both of them. Note that a solution to converge to the exact ground-state energy is to increase the number of measures to reduce the fluctuations due to the sampling noise and resolve the small spectral gap.
}

Therefore, starting from an initial single-mode state, cat-like states seem difficult to realize by the BS-Kerr ansatz when sampling noise is present.
However, much better accuracy is reached when the two-mode initial state is used instead, thus suggesting that starting from a configuration with photons spread in between the modes can improve the results of the VQE
in the strongly correlated regime ($\Lambda \geq 5$ here).
In summary, it seems from Fig.~\ref{fig:fidelity_map}
that -- within the BS-Kerr ansatz -- a weakly-correlated state is more easily encoded when starting from a single-mode state rather than a two-mode one, and vice versa for a strongly-correlated state.
{Note that better results could possibly be reached by using other noise-adapted optimizers rather than CMA-ES used in this work,
see Refs.~\cite{lavrijsen2020classical,sung2020using} and references therein.
If the optimization becomes a problem when tackling larger systems,
one can decompose it into smaller pieces that are iteratively solved by small sized circuits, as proposed by Fujii {\it et al.} with the so-called deep VQE based on the divide and conquer approach~\cite{fujii2020deep}.}

\section{Conclusions}\label{sec:conclusion}

In this work, we tackled the problem of encoding strongly correlated many-boson ground-state wavefunctions on a photonic quantum device. {A particular attention was paid to the expressibility and the resource efficiency of the circuits for different regimes of many-body interaction.}
For this, we designed two different quantum photonic-based ansatze
and simulated a variational quantum algorithm (VQA) using the ground-state infidelity as a cost function.
Both ansatze, the beam-splitter-Kerr (BS-Kerr) and the interferometer-Kerr, are based on layers of beam-splitter and Kerr gates, with additional rotation gates for the latter.
We showed their ability to encode the ground state of the attractive BH model in all correlation regimes, for different number of bosons and initial states.
While the BS-Kerr ansatz has shown to be the most efficient in terms of total number of gates and parameters, it might not be optimal for conducting a true experiment on a real photonic quantum device, as it contains many Kerr gates that are difficult to implement in practice.
In contrast, the interferometer-Kerr ansatz was designed to reduce the number of Kerr gates (or the number of layers required to achieve a given ground-state fidelity) at the expense of more rotation and beam-splitter gates and more parameters per layer, and is therefore more appropriate for realistic applications.
Although taking the fidelity as a criteria is theoretically sound to test the encoding ability of our ansatze,
in practice the cost function is the ground-state energy measured within the variational quantum eigensolver (VQE) algorithm.
As a last investigation, we performed realistic VQE simulations using the BS-Kerr ansatz, with and without sampling noise, and proposed
a photon-counting protocol to measure the ground-state energy of the BH Hamiltonian as it would be done in a real experiment.

{
As further suggestions for future work, one could include noise and losses in realistic simulations of actual photonic circuitry, as well as time-frequency effects which affect non-linear optical elements such as squeezers and, presumably, Kerr media. Moreover, although the ansatze proposed in our work were designed to solve the BH model, they could in principle be used for a larger range of applications, such as the absorption of Helium-4 on graphite~\cite{kwon20124,gordillo2009he}.}

\section*{Acknowledgments} 
SY and BS sincerely thank Eleanor Scerri, Xavi Bonet-Monroig and Vincent Pouthier for fruitful discussions.
SY and BS acknowledge support from the Netherlands Organization for Scientific Research (NWO/OCW). SY also acknowledges the Interdisciplinary Thematic Institute ITI-CSC
via the IdEx Unistra (ANR-10-IDEX-0002) within the program Investissement d’Avenir.
VD acknowledges support  by  the  Dutch  Research  Council (NWO/OCW),  as  part  of  the  Quantum  Software  Consortium program (project number 024.003.037).



\appendix

\section{Exact encoding of the BH ground state using only beam-splitters in the non-interacting limit}\label{app:demo}

We demonstrate here that a single layer of the BS-Kerr ansatz, without the Kerr gates (so only beam-splitters) can exactly encode the ground state of the BH model for the three newtorks considered in our study (i.e.~with $N_S=2$, 3 and 4 sites) in the non-interacting limit $\Lambda \rightarrow 0$.
Note that this demonstration works only for initial single-mode states such as $\ket{N_B,0}$, $\ket{N_B,0,0}$ and $\ket{N_B,0,0,0}$ for the dimer, three- and four-site BH model, respectively.

Let us start with the dimer, for which the exact ground state reads
\begin{equation}\label{eq:exact_GS_dimer}
    \ket{\Psi_0} = \frac{1}{\sqrt{2N_B!}} [  a_1^\dagger +  a_2^\dagger ]^{N_B}\ket{0,0}.
\end{equation}
The action of a single layer of beam-splitter gates on an initial single-mode state results in
\begin{equation}
\begin{split}
        B_{1,2}(\theta,0) \ket{N_B,0} &=\\ 
    \frac{1}{\sqrt{N_B!}} &[ \cos(\theta)a_1^\dagger + \sin(\theta) a_2^\dagger ]^{N_B}\ket{0,0}
\end{split}
\end{equation}
which provides an exact encoding of the ground state [Eq.~(\ref{eq:exact_GS_dimer})] if we fix the beam-splitter transmittivity parameter to $\theta = \pi/4$. 
Turning to the three-site model, its exact ground state is given by
\begin{equation}\label{eq:exact_GS_trimer}
    \ket{\Psi_0} = \frac{1}{\sqrt{3N_B!}} [  a_1^\dagger +  a_2^\dagger +  a_3^\dagger]^{N_B}\ket{0,0,0}.
\end{equation}
applying the stair structure of the beam-splitters transform the initial single-mode state as follows,
\begin{equation}
\begin{split}
          &B_{2,3}(\theta',0) B_{1,2}(\theta,0) \ket{N_B,0,0} = \\
          &\frac{1}{\sqrt{N_B!}} \Big[ \cos(\theta)a_1^\dagger + \sin(\theta) \cos(\theta') a_2^\dagger \\
          &+ \sin(\theta) \sin(\theta')a_3^\dagger   \Big]^{N_B}\ket{0,0,0}  ,
\end{split}
\end{equation}
thus realizing an exact encoding of the ground state [Eq.~(\ref{eq:exact_GS_trimer})] when we fix the beam-splitter's parameters such that $\theta = \arccos(1/\sqrt{3})$ and $\theta'=\pi/4$. 
Finally, the exact ground state of the four-site model reads
\begin{equation}\label{eq:exact_GS_quadrimer}
    \ket{\Psi_0} = \frac{1}{\sqrt{4N_B!}} [  a_1^\dagger +  a_2^\dagger +  a_3^\dagger+  a_4^\dagger]^{N_B}\ket{0,0,0,0}.
\end{equation}
Similarly as for the BH dimer and three-site models, we apply a single layer of the BS-Kerr ansatz (without the Kerr) on the single-mode initial state, thus leading to
\begin{equation}
\begin{split}
&B_{3,4}(\theta'',0)  B_{2,3}(\theta',0) B_{1,2}(\theta,0) \ket{N_B,0,0,0} =\\
&\frac{1}{\sqrt{N_B!}} \Big[ \cos(\theta)a_1^\dagger + \sin(\theta) \cos(\theta') a_2^\dagger\\
&+  \sin(\theta)\sin(\theta') \cos(\theta'')a_3^\dagger\\
&+\sin(\theta)\sin(\theta')\sin(\theta'')a_4^\dagger\big)\big)   \Big]^{N_B}\ket{0,0,0,0}.
\end{split}
\end{equation}
Fixing the beam-splitter's parameters to $\theta=\pi/3$, $\theta'=\arccos(1/\sqrt{3})$ and $\theta''=\pi/4$ generates an exact representation of the exact ground state in Eq.(\ref{eq:exact_GS_quadrimer}).

\section{ Action of the BS-Kerr ansatz on the BH dimer with 2 bosons } \label{appendix:illustration}

In this section, we illustrate the functioning of the BS-Kerr ansatz to solve the ground state of the Bose-Hubbard model. We focus  on the simplest case possible: a dimer containing two bosons (\textit{i.e. } $N_S=N_B=2$).  As a reference, we consider the exact form of the ground state for an intermediate interaction regime $\Lambda = 6$, which is given by 
\begin{align}
  \ket{\Psi_0} =  \frac{1}{\sqrt{5}} \ket{11} +  \sqrt{\frac{2}{5}} \big( \ket{20} + \ket{02} \big).
  \label{eq:exact_state}
\end{align}
For the illustration, let us consider an initial homogeneous distribution of the bosons in the two modes, i.e. an initial state $\ket{11}$. 
Starting from this, we will demonstrate that a single layer of the BS-Kerr ansatz is sufficient to exactly encode the ground state of the system.
The associated state generated by the circuit is here
\begin{equation}
    \ket{\Psi (\vec{\theta})} = K_1(\theta_1) K_2(\theta_2)  B(\theta)  \ket{11}
\end{equation}
Focusing first on the BS-gate's action, we can show (using Eq.~\ref{eq:BS_action}) that
\begin{align}
    \begin{split}
      B(\theta)  \ket{11} = \cos(2\theta) \ket{11} + \frac{\sin(2\theta)}{\sqrt{2}}  \big(\ket{02} - \ket{20} \big)
  \end{split}.
\end{align}
This result clearly illustrates the role of the BS gate: it spreads the bosonic wavefunction over the three different accessible Fock states of the system. 
Applying Kerr gates on this quantum superposition of Fock state introduces non-linear phases on each independent Fock state such as
\begin{equation}
    \begin{split}
        \ket{\Psi (\vec{\theta})} &= K_1(\theta_1) K_2(\theta_2)  B(\theta)  \ket{11}  \\
                    &= e^{i(\theta_1+\theta_2)}\cos(2\theta) \ket{11} \\
                    &+  \frac{\sin(2\theta)}{\sqrt{2}}  \big( e^{i4\theta_2}\ket{02} - e^{i4\theta_1}\ket{20}  \big).
    \end{split}
    \label{eq:ansatz_2modes}
\end{equation}
In this case, encoding the exact system's ground state with the three gates involved means solving the set of coupled non-linear equations to match the best the coefficients in \ref{eq:ansatz_2modes} and \ref{eq:exact_state}. 
An exact solution of the problem is given by $\theta_1 = 3\pi/8$, $\theta_2 = \pi/8$ and $\theta = \arccos(1/\sqrt{5}) /2$. 
Note that replacing Kerr-gates with simple rotation-gates makes the exact encoding of the ground state impossible.

\phantomsection
\bibliographystyle{unsrtdin}
\bibliography{biblio}


\end{document}